\documentclass[aps,prb,showpacs,citeautoscript,twocolumn,superscriptaddress]{revtex4-1}

\usepackage{graphicx}
\usepackage{amsmath}
\usepackage{amssymb}
\usepackage{color}
\usepackage{subfigure}
\usepackage{braket}

\newcommand{\bea}{\begin{eqnarray*}}
\newcommand{\eea}{\end{eqnarray*}}
\newcommand{\bne}{\begin{equation*}}
\newcommand{\ede}{\end{equation*}}

\newcommand{\bnen}{\begin{equation}}
\newcommand{\eden}{\end{equation}}
\newcommand{\bean}{\begin{eqnarray}}
\newcommand{\eean}{\end{eqnarray}}
\newcommand{\bnsn}{\begin{subequations}}
\newcommand{\edsn}{\end{subequations}}

\newcommand{\bna}{\begin{array}}
\newcommand{\eda}{\end{array}}
\newcommand{\bnm}{\begin{enumerate}}
\newcommand{\edm}{\end{enumerate}}
\newcommand{\bni}{\begin{itemize}}
\newcommand{\edi}{\end{itemize}}

\renewcommand{\vec}[1]{\text{\boldmath{$ #1 $}}}

\begin{document}

\title{Valley-enhanced fast
relaxation of 
gate-controlled donor qubits in silicon}

\author{P\'eter Boross}
\affiliation{Institute of Physics, E\"otv\"os University, Budapest, Hungary}

\author{G\'abor Sz\'echenyi}
\affiliation{Institute of Physics, E\"otv\"os University, Budapest, Hungary}

\author{Andr\'as P\'alyi}
\affiliation{
Department of Physics and 
MTA-BME Condensed Matter Research Group,
Budapest University of Technology and Economics, Budapest, Hungary}

\date{\today}

\begin{abstract}
Gate control of donor electrons near interfaces is
a generic ingredient of donor-based quantum computing. 
Here, we address the question: 
how is the phonon-assisted qubit relaxation time $T_1$
affected as the electron is shuttled between the donor and the 
interface?
We focus on the example of the 
`flip-flop qubit' [Tosi {\it et al.}, arXiv:1509.08538v1],
defined as a combination of the
nuclear and electronic states of a phosphorous donor
in silicon, 
promising
fast electrical control and
long dephasing times
when the electron is halfway
between the donor and the interface.
We theoretically describe
orbital relaxation,
flip-flop relaxation,
and 
electron spin relaxation. 
We estimate that the flip-flop 
qubit relaxation time can be
of the order of $100 \, \mu\text{s}$,
8 orders of magnitude shorter than the value for 
an on-donor electron in bulk silicon,
and a few orders of magnitude shorter (longer) than 
the predicted inhomogeneous dephasing time (gate times).
All three relaxation processes are boosted
by   
(i) the nontrivial valley structure of the electron-phonon
interaction, and 
(ii) the different valley compositions of the involved electronic states. 
\end{abstract}

\maketitle

\section{Introduction}

Donor-based spin qubits in 
silicon\cite{Kane_Nature98, Morton_Si_QC_QmLim_Nat11,Zwanenburg_SiQmEl_RMP13}  (Si)
are promising building blocks
for quantum information processing schemes, 
mainly due to  qubit lifetimes that are prolonged by the
weakness of spin-orbit and hyperfine interactions in this material\cite{Feher_PR59, FeherGere_PR59, Roth_PR60, Hasegawa_PR60,
Tahan2002,Tyryshkin_PRB03,Tyryshkin_JPC06,Morello_1glShot_Nature10, Tyryshkin_IsoPureSiDnr_T2Secs_11, Tahan2014}. 
Recent important experimental achievements of the field 
include initialization, coherent control and readout
of electronic and nuclear spins of individual phosphorous (P) donors\cite{Morello_1glShot_Nature10, Pla-electronspin, Pla-nuclearspin},
as well as increasing 
qubit lifetimes \cite{Tyryshkin_PRB03,Tyryshkin_JPC06,Tyryshkin_IsoPureSiDnr_T2Secs_11,Muhonen_Store_14,Itoh_MRS} by using isotopically purified
samples with strongly increased abundance of the 
nuclear-spin-free Si-28 isotope.

\begin{figure}
\begin{center}
\includegraphics[width=1\columnwidth]{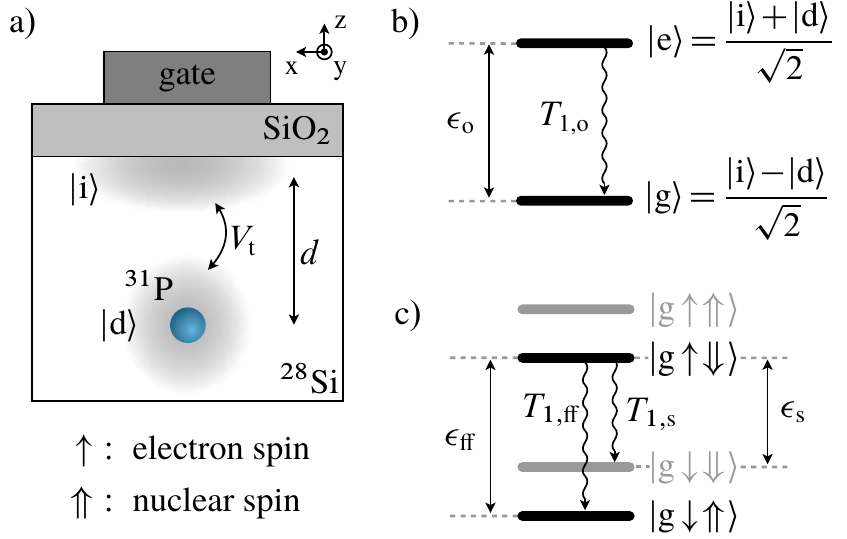}
\end{center}
\caption{(Color online) Flip-flop qubit: setup and relaxation processes.
(a) A donor is placed in the vicinity of a Si/SiO$_2$ interface,
so that its electron can be moved between the
interface ($\ket{\text{i}}$) and the donor ($\ket{\text{d}}$) by the voltage applied 
to the gate electrode.
$V_\text{t}$: tunnel amplitude between $\ket{\text{i}}$ and $\ket{\text{d}}$;
$d$: distance between the charge centers associated to 
$\ket{\text{i}}$ and $\ket{\text{d}}$. 
(b) Charge qubit basis states and 
orbital relaxation ($T_\text{1,o}$)
at the ionization point, $\epsilon_\text{o} = V_\text{t}$. 
(c) Energy diagram showing the combined states of the 
electronic and nuclear spins of the donor. 
The flip-flop qubit basis states are highlighted as thick
black lines, their energy splitting is $\epsilon_\text{ff}$.
Flip-flop relaxation ($T_\text{1,ff}$) and 
electron spin relaxation ($T_\text{1,s}$) both lead to information loss. 
}
\label{fig:setup}
\end{figure}

A ubiquitous ingredient of donor-based quantum-information 
processing schemes is to use electrical gates to control
the wave function of the donor electron (see Fig.~\ref{fig:setup}a).
That often means that the electron
is shuttled between the donor and a nearby interface\cite{Kane_Nature98,Vrijen_PRA00,Calderon_quantumcontrol,Calderon_PRB08,Lansbergen_donorcontrol,Rahman-orbitalstark,Baena,Laucht,Tosi,Urdampilleta,HarveyCollard}.
For example, in the Kane proposal\cite{Kane_Nature98},
gate control is suggested 
to tune the hyperfine interaction strength and
to allow for exchange-based two-qubit operations.
Here, we address the following question: 
how does the phonon-assisted qubit relaxation time $T_1$
depend on the location of the electron,
as it is placed in an intermediate position between the donor
and the interface? 
We focus on the example of the recently proposed\cite{Tosi}
\emph{flip-flop qubit} (see Fig.~\ref{fig:setup}c);
it is defined as a combination of the
nuclear and electronic states of a phosphorous donor
in silicon, 
and it is expected to allow for 
fast electrical control and
long dephasing times,
when the gate-induced electric fields locate the
electron halfway between the donor and the interface.

\subsection{Flip-flop qubit}

Naturally, most of the coherent-control experiments with 
donor-based spin qubits are performed using ac magnetic fields
in the spirit of paramagnetic resonance. 
However, for a number of practical reasons, it can be advantageous 
to substitute the magnetic excitation with electrical driving, 
which is possible if
a sufficiently 
strong interaction exists between the spin qubit and 
electric fields.
On a single-qubit level, such an interaction allows local 
control via ac gate-voltage pulses\cite{Golovach_EDSR_PRB06,Flindt,Nowack_Science07}, 
and dispersive non-demolition
readout via probing a nearby electromagnetic resonator\cite{Blais}.
It also enables two-qubit
operations, either via electric dipole-dipole 
interaction\cite{Flindt,Trif_spinspincoupling,Tosi,Salfi_acceptorqubit}, 
or via an electromagnetic resonator that mediates
interaction between the qubits\cite{Blais}.
These two-qubit gates, in contrast to the 
exchange-based gate,
should be robust against donor placement uncertainties\cite{Cullis,Kane_Nature98,Koiller_PRL01}.

The flip-flop qubit\cite{Tosi}
is expected to interact strongly with 
electric fields, and therefore has the potential to 
realize the desired features outlined above.
The qubit is encoded in the composite system of the
electronic and nuclear spins of a P donor, such that the qubit
basis states are given by the two anti-aligned spin configurations
$\uparrow \Downarrow$ and $\downarrow \Uparrow$,
where the first (second) arrow represents the electronic (nuclear)
spin.
Importantly, the flip-flop terms of the hyperfine interaction
between the electronic and nuclear spins couple the
two qubit basis states.
As a consequence, an ac electric field can drive coherent Rabi 
oscillations of the qubit: the field shakes the electronic wave
function, thereby modulates the hyperfine coupling strength, 
which is in turn felt by the qubit as an ac Hamiltonian
matrix element 
that couples the basis states. 

This interaction between the flip-flop qubit and  electric fields
can be 
strongly enhanced in the configuration shown in Fig.~\ref{fig:setup}a. 
Here, the donor is placed in the vicinity of an interface
between silicon and a barrier material (e.g., SiO$_2$).
If the charge center of the electron 
is approximately halfway between the donor ion
and the interface (\emph{ionization point}), then
the coupling between the qubit and
electric fields is maximized.
A further advantage of such a setting is the existence
of dephasing sweet spots in the space of the 
control parameters, including 
second-order clock-transition points
where  both the first and second derivatives of the 
qubit's Larmor frequency with respect to the dc electric field
are zero.
Tuning the system to such a sweet spot might 
result in exceptionally strong resilience against 
electrically-induced dephasing.

\subsection{This work}

In this work, we theoretically describe
phonon-mediated relaxation of the flip-flop qubit,
and determine the corresponding relaxation time $T_\text{1,ff}$
(see Fig.~\ref{fig:setup}c and section \ref{sec:ffq}).
Reference \onlinecite{Pines} estimated 
a very long low-temperature 
relaxation time of $T_\text{1,ff} \sim 10^4$ s 
for a P donor in bulk silicon, set by deformation-induced
changes of the 
effective mass and the dielectric constant.
(Phonon-mediated spin relaxation processes involving 
nuclear-spin \emph{ensembles} are treated, e.g., in
Refs.~\onlinecite{Abragam,Khaetskii-physicae,Erlingsson_2002_hyperfine}.)
In contrast, here we describe a
deformation-potential mechanism
that is particularly strong in the proposed working point
of the flip-flop qubit, 
when the electron is at the ionization point, 
and leads to a
characteristic $T_\textrm{1,ff} \sim 100\, \mu$s.
This time scale 
is approximately 8 orders of magnitude shorter than the 
prediction of Ref.~\onlinecite{Pines},
and a few orders of magnitude shorter (longer) than the 
predicted\cite{Tosi} inhomogeneous dephasing time (gate times) of the
flip-flop qubit.

The reason for the relatively fast relaxation is twofold. 
First, the flip-flop qubit is designed to strongly interact
with electric fields at its working point, and that is achieved via 
hyperfine-induced mixing of the ground-state orbital 
with a low-lying excited orbital\cite{Tosi}
($\ket{\text{g}}$ and $\ket{\text{e}}$, to be introduced below).
The same low-lying excited orbital also
provides strong interaction between the flip-flop qubit and
phonon-induced deformation
potentials,
leading to relatively fast qubit relaxation.
Second, 
we show that the relaxation process is \emph{valley-enhanced},
where \emph{valley} refers to the 6 conduction-band minima
of the electronic band structure of silicon. 
In particular, the relaxation is boosted by 
the nontrivial valley-related features of 
the electron-phonon interaction and the involved electronic states.

We also characterize orbital relaxation,
that is, relaxation of the charge qubit
($T_{1,\text{o}}$ in Fig.~\ref{fig:setup}b).
Since orbital relaxation is
conceptually simpler than the flip-flop relaxation, we 
start with the case of orbital relaxation in section \ref{sec:charge},
and use it 
to introduce and illustrate the key ingredients of the valley-enhanced 
mechanism that  governs all
the three processes we consider (orbital, flip-flop, and electron
spin relaxation).
Finally, in section \ref{sec:spin},
we describe
electron spin relaxation\cite{Feher_PR59, FeherGere_PR59, Roth_PR60, Hasegawa_PR60, Tahan2002,Morello_1glShot_Nature10,Tahan2014} from the
excited state of the flip-flop qubit
($T_\text{1,s}$ in Fig.~\ref{fig:setup}c): this process 
is also relevant for the functionality of the flip-flop qubit, as
it leads to leakage from the qubit subspace.


\section{The flip-flop qubit and its model
Hamiltonian}
\label{sec:ffqintro}

Here, based on Ref.~\onlinecite{Tosi},  we discuss the setup
in which the flip-flop qubit is envisioned, 
a simple 8-dimensional model Hamiltonian that
captures the essential ingredients of the setup, 
and the 2-dimensional flip-flop qubit subspace.
We note that in Ref.~\onlinecite{Tosi}, this  8-dimensional
Hamiltonian was found to 
reliably reproduce various physical quantities obtained from
atomistic tight-binding simulations. 
This fact promotes this
 model to a trustable starting point for exploring the
relaxation mechanisms of the flip-flop qubit.

In the absence of gate-induced electric fields, the donor electron
is localized at the donor site, occupying the 
ground-state donor orbital $\ket{\text{d}}$ (see Fig.~\ref{fig:setup}a). 
A voltage applied on the gate electrode induces an electric 
field along the z axis, and hence can pull the electron 
to the vicinity of
the silicon-barrier interface, where it occupies the orbital
state $\ket{\text{i}}$.
This two-orbital \emph{charge qubit} degree of freedom is
described by the Pauli matrices 
$\sigma_\text{x}$, $\sigma_\text{y}$, $\sigma_\text{z}$, 
where, e.g., $\sigma_\text{z} = \ket{\text{i}}\bra{\text{i}} - \ket{\text{d}} \bra{\text{d}}$.
By continuously changing the gate-induced electric field
$E_\text{z}$, the electron is  continuously moved between the 
two localized orbitals; the corresponding Hamiltonian reads
\bean
\label{eq:chargeham}
H_\text{o}=\frac{V_\text{t}}{2}\sigma_\text{x}-\frac{e(E_\text{z}-E_\text{z}^0)d}{2}\sigma_\text{z},
\eean
where $V_\text{t}$ is the tunnel amplitude between the
orbital states $\ket{\text{i}}$ and $\ket{\text{d}}$,
and $E_\text{z}^0$ is the value of the gate-induced electric field along z
where the stationary electron charge is equally
distributed among $\ket{\text{i}}$ and $\ket{\text{d}}$.
The splitting between the energy eigenvalues of 
$H_\text{o}$ is
\bean
\epsilon_\text{o} = \sqrt{V_\text{t}^2 + 
\left[ e (E_\text{z} - E_\text{z}^0) d\right]^2}.
\eean

An external homogeneous magnetic field
introduces Zeeman splittings for both the electron and the nuclear
spin of the donor.
For simplicity, for the moment 
we assume isotropic and location-independent
$g$-tensors, yielding the following electronic and nuclear
Zeeman Hamiltonians, respectively:
\bean
H_{B,e} = h \gamma_\text{e} \vec B \vec S,
\eean
\bean
H_{B,n} = h \gamma_\text{n} \vec B \vec I.
\eean

If the electron is located on the donor, then 
its spin $\vec S$
interacts with the nuclear spin $\vec I$ of the donor. 
Hence the hyperfine interaction is described by
the following Hamiltonian: 
\bean
H_\text{hf}=A\left(\frac{1-\sigma_\text{z}}{2}\right)\vec{S}\vec{I}.
\eean
Here, both $\vec S$ and $\vec I$ are represented by $1/2$ times
the vector of Pauli matrices.
We introduce the \emph{secular} $H_\text{hf,sec}$ and 
\emph{non-secular} or \emph{flip-flop} part
$H_\text{hf,ff} = H_\text{hf} - H_\text{hf,sec}$
of the hyperfine Hamiltonian, where the former is defined as
\bean
H_\text{hf,sec} =
A\left(\frac{1-\sigma_\text{z}}{2}\right)
\left(
	\vec{S} \cdot \frac{\vec B}{B}
\right)
\left(
	\vec{I} \cdot \frac{\vec B}{B}
\right).
\eean
That is, $H_\text{hf,sec}$ incorporates 
spin components that are 
parallel to the external magnetic field,
whereas $H_\text{hf,ff}$ incorporates the flip-flop terms. 

The energy eigenstates of the $8 \times 8$ Hamiltonian
$H_\text{sec}=H_\text{o}+H_{B,n} + H_{B,e}+H_\text{hf,sec}$ 
are direct products of the energy eigenstates $\ket{\text{g}}$
and $\ket{\text{e}}$ of 
$H_\text{o}$ and electron ($\uparrow$, $\downarrow$)
and nuclear ($\Uparrow$, $\Downarrow$) spin states
pointing along the external magnetic field.
These states will be labelled by the above
quantum numbers and denoted as, e.g., 
$\ket{\text{g}\!\downarrow \Uparrow}_0$, 
and we will call them the \emph{unperturbed energy eigenstates}.

If the spectral gaps of $H_\text{sec}$
are much larger than the energy scale $A/4$ characterizing
$H_\text{hf,ff}$,
then the latter remains a perturbation,
and the energy eigenstates of the full Hamiltonian
$H = H_\text{sec} + H_\text{hf,ff}$ are approximately
direct products as above, hence can be labelled with
the same quantum numbers, and will be denoted
as, e.g., $\ket{\text{g}\!\downarrow \Uparrow}$.
Using this notation,
the basis states of the \emph{flip-flop qubit} are
$\ket{\text{g}\!\downarrow \Uparrow}$
and $\ket{\text{g}\!\uparrow \Downarrow}$.
An example parameter set where the above conditions are met, 
and which was studied extensively in Ref.~\onlinecite{Tosi}, is 
shown in Table \ref{tab:params}.
The level diagram consisting of the four energy eigenstates
of $H$ associated to the ground-state orbital manifold
is depicted in Fig.~\ref{fig:setup}c;
there, the flip-flop qubit basis states,
having an energy separation of 
$\epsilon_\text{ff} \approx h \gamma_\text{e} B$, are highlighted 
as bold black lines.

\begingroup
\squeezetable
\begin{table}
\caption{\label{tab:params}
Parameter values.
Remarks:
(1)
Working-point parameters are taken from 
Ref.~\onlinecite{Tosi}, except the value 
of $E_\text{z}-E_\text{z}^0$, which we set to
zero for simplicity.
(2) 
Having no estimate for the spin-dependent tunnel matrix element
$V_\text{s}$, the quoted value is an arbitrary choice. 
}
\begin{tabular}{l  c  c}
Parameter & Notation & Value \\
\hline \hline
Material-specific parameters \\ \hline
Mass density of silicon & $\rho$ & 2330 kg/m$^3$ \\
Uniaxial deformation potential & $\Xi_u$ & $8.77$ eV \\
Longitudinal sound velocity & $v_\text{L}$ & 9330 m/s \\
Transverse sound velocity & $v_\text{T}$ & 5420 m/s \\
Hyperfine interaction strength for Si:P & $A/h$ & 117 MHz \\
Gyromagnetic ratio of electron spin & $\gamma_\text{e}$ & 27.97 GHz/T \\
Gyromagnetic ratio of P nuclear spin & $\gamma_\text{n}$ & 17.23 MHz/T \\
\hline
Working-point parameters \\
\hline
Magnetic field & $B$ & 0.2 T  \\
Electric-field detuning from ionizaton point & 
$E_\text{z} - E_\text{z}^0$ & 0 V/m \\
Donor-interface hopping amplitude & $V_\text{t}/h$ & 5.91 GHz \\
Donor-interface center-of-charge distance & $d$ & 15 nm \\
\hline
Further parameters \\
\hline
Spin-dependent tunnel matrix element & $V_\text{s}/h$ & 10 MHz \\
Perpendicular $g$-tensor anisotropy\cite{Rahman_gfactor} 
& $\Delta_\gamma^\perp$ & -0.2\% 
\\
Parallel $g$-tensor anisotropy\cite{Rahman_gfactor} & $\Delta_\gamma^\parallel$ & 0.7\%
\\
Amplitude of ac electric field & $E_\text{ac}$ & 32 V/m
\end{tabular}
\end{table}
\endgroup

\section{Orbital relaxation}
\label{sec:charge}

First, we characterize the phonon-emission-mediated orbital 
relaxation, that is, 
relaxation from the excited state $\ket{\text{e}}$
of the charge qubit
to its ground state $\ket{\text{g}}$, and calculate
the corresponding relaxation time $T_\text{1,o}$,
see Fig.~\ref{fig:setup}b. 
We disregard the spin degrees of freedom for simplicity;
the charge qubit is described by the $2\times 2$
Hamiltonian $H_\text{o}$ of Eq.~\eqref{eq:chargeham},
and the eigenstates 
$\ket{\text{g}}$ and $\ket{\text{e}}$
of $H_\text{o}$ are called
the \emph{charge qubit basis states}. 
The valley-enhanced, deformation-potential-induced 
relaxation mechanism we describe here, 
as well as the structure of the calculation itself,  is  
easily translated to treat the 
flip-flop relaxation and electron spin relaxation processes, 
which will be discussed in the subsequent sections. 

\subsection{Preliminaries}

To account for the phonons and the electron-phonon interaction,
we use a bulk-type description, neglecting any effects arising from
inhomogeneities in the nanostructure. 

In the experimentally relevant range of parameters, the
charge-qubit energy splitting is resonant with low-energy
long-wavelength acoustic phonons. 
Hence only those are considered here.
Their dispersion relations are  
assumed to be linear and characterized by the
sound velocities $v_\lambda$, where
$\lambda \in$ (L,T1,T2) is the polarization index
and L (T) refers to longitudinal (transverse). 

We focus on the case of zero temperature
and use the corresponding Fermi's Golden Rule
to evaluate the qubit relaxation time:
\bean
\label{eq:fgr}
\frac 1 {T_\text{1,o}} = 
\frac{2\pi}{\hbar} 
\sum_{\vec q,\lambda}
\left| 
\braket{\text{g}, \vec q \lambda 
| H_\text{eph} |  \text{e}, 0}
\right|^2 
\delta(\epsilon_\text{o} - \hbar v_\lambda q).
\eean
Here, bras and kets represent  joint states of the
composite electron-phonon system,  
$0$ denotes the vacuum of phonons,
and 
$\vec q$ ($\lambda$) is the wave number 
(polarization index) of the emitted phonon.

The mechanism we describe is based on the deformation-potential
electron-phonon interaction, which we treat via the silicon-specific 
Herring-Vogt
Hamiltonian\cite{Herring,Yu_Cardona}:
\bean
\label{eq:herringvogt}
H_\text{eph} = 
\Xi_u
\left(
\bna{cccccc}
\varepsilon_{xx} & 0 & 0 & 0 & 0 & 0 \\
0 & \varepsilon_{xx} & 0 & 0 & 0 & 0 \\
0 & 0 & \varepsilon_{yy} & 0 & 0 & 0 \\
0 & 0 & 0 & \varepsilon_{yy} & 0 & 0 \\
0 & 0 & 0 & 0 & \varepsilon_{zz} & 0 \\
0 & 0 & 0 & 0 & 0 & \varepsilon_{zz}
\eda
\right),
\eean
where the $6\times 6$ matrix structure corresponds to 
\emph{valley space}, that is, 
the 6 envelope functions associated to the 6 conduction-band
valleys of silicon, 
denoted and ordered as $(x,\bar x, y, \bar y, z, \bar z)$.
In Eq.~\eqref{eq:herringvogt}, 
$\Xi_u$ is the uniaxial deformation potential and
$\varepsilon$ is the 
strain tensor.
Note that in addition to the right hand side of Eq.~\eqref{eq:herringvogt}, 
the Herring-Vogt Hamiltonian incorporates
a conventional, valley-independent 
deformation-potential term,
$\Xi_d \text{Tr}(\varepsilon) 1_{6\times 6}$, 
where $\Xi_d$ is the dilational deformation potential, 
$\text{Tr}(\varepsilon)$ 
is the deformation-induced relative volume change, and
$1_{6\times 6}$ is the $6\times 6$ unit matrix;
however, we disregard that term here as 
(i) it does not contribute
to the valley-enhanced mechanism to be described here, and 
(ii) its contributions to the relaxation rates obtained here
are much smaller than those of the uniaxial deformation potential term.

The diagonal elements of the strain tensor,
that is, the elements that determine  $H_\text{eph}$
via Eq.~\eqref{eq:herringvogt},
read
\bean
\label{eq:strainunexpanded}
\varepsilon_{jj} =
i \sqrt{\frac{\hbar}{2 \rho V}} 
\sum_{\vec q,\lambda} 
\frac{e_{\vec q \lambda j} q_j}{\sqrt{ v_\lambda  q}}
e^{i \vec q \cdot \vec r}
\left(
	a_{\vec q,\lambda} + a^\dag_{-\vec q,\lambda}
\right).
\eean
Here, $j \in \{\text{x},\text{y},\text{z}\}$,
$\rho$ is the mass density of silicon,
$V$ is the sample volume and 
$\vec e_{\vec q \lambda}$ is the polarization vector 
of the phonon mode with wave number $\vec q$ 
and polarization index $\lambda$.
For the setup we consider, the wavelength of 
the phonon emitted by the qubit is much longer than
the spatial size of the qubit itself. 
Therefore, the plane-wave factor in Eq.~\eqref{eq:strainunexpanded}
can be approximated as
\bean
e^{i\vec q \cdot \vec r} \approx 1; 
\label{eq:homogeneousdeformation}
\eean
this corresponds to a homogeneous deformation, 
and as we will show, such a homogeneous 
deformation is sufficient to induce the 
described relaxation processes.

To obtain $T_\text{1,o}$ via Fermi's Golden Rule \eqref{eq:fgr},
we need to provide the envelope-function
representation of the localized charge states $\ket{\text{i}}$ and
$\ket{\text{d}}$. 
For the purpose of obtaining the order of magnitude 
and the parameter dependence of the relaxation rates,
it is sufficient to use  simple
`perfectly localized' envelope functions,
dressed by the appropriate
valley compositions\cite{Kohn_donorstates,Baena,Zwanenburg_SiQmEl_RMP13}.
The interface state $\ket{\text{i}}$
resembles that of a planar quantum-dot 
ground state pushed toward the barrier
by the gate-induced electric field, hence its wave function
resides in the $z$ and $\bar z$ valleys,
evenly distributed. 
The donor state $\ket{\text{d}}$, on the other hand, is
evenly distributed in all the 6 valleys. 
Using these considerations, we represent the two
localized charge states as 
\begin{subequations}
\label{eq:valleycompositions}
\bean
\label{eq:valleyi}
\braket{\vec r | \text{i} } &=&  
 \sqrt{\delta(\vec r- \vec r_\text{i})}
\frac{1}{\sqrt 2}
(0,0,0,0,e^{i\phi_{z}},e^{i\phi_{\bar{z}}}), \\
\label{eq:valleyd}
\braket{\vec r | \text{d} } &=& 
\sqrt{\delta(\vec r)}
\frac 1 {\sqrt{6}}
(1,1,1,1,1,1),
\eean
\end{subequations}
where $\delta(\vec r)$ is the three-dimensional 
Dirac delta, 
the donor position is chosen as the origin of the 
reference frame, $\vec r_\text{i}$
is the center of charge of the orbital $\ket{\text{i}}$, 
and the phases $\phi_z$ and $\phi_{\bar{z}}$ are between
0 and $2\pi$, but their actual values turn out to be irrelevant.
In Eq.~\eqref{eq:valleycompositions}, the Dirac delta
is a strongly simplified representation of the envelope functions 
associated to the valleys.
We emphasize that a more realistic representation, e.g., 
using Kohn-Luttinger\cite{Kohn_donorstates} envelope functions
for the donor orbital $\ket{\text{d}}$, would only lead to 
minor quantitative corrections of our results.

Before evaluating the orbital relaxation time, it is instructive
to restrict the electron-phonon interaction Hamiltonian $H_\text{eph}$
to the charge-qubit Hilbert space:
\bean
\label{eq:ephc}
H_\text{eph,o} = P H_\text{eph} P
=
\Xi_u
\frac i 6 \sqrt{\frac{\hbar}{2\rho V}}
\Sigma_\text{z} \sigma_\text{z},
\eean
where $P= \ket{\text{i}}\bra{\text{i}} + \ket{\text{d}}\bra{\text{d}}$,
\begin{align}
\label{eq:ephcSigma}
\Sigma_\text{z} &=& 
\sum_{\vec q, \lambda} 
\frac{
\left(
	- e_{\vec q \lambda x} q_x 
	- e_{\vec q \lambda y} q_y +
	2 e_{\vec q \lambda z} q_z
\right)
}{
\sqrt{v_\lambda q}}
\left(
	a_{\vec q,\lambda} + a^\dag_{-\vec q,\lambda}
\right).
\end{align}
Here 
we used Eqs. \eqref{eq:herringvogt}, \eqref{eq:strainunexpanded},
\eqref{eq:homogeneousdeformation} and \eqref{eq:valleycompositions},
and from Eq.~\eqref{eq:ephc} we omitted an irrelevant term 
proportional to the unit matrix $\sigma_0$.
Remarkably, $H_\text{eph,o}$ is proportional to 
$\sigma_\text{z}$, which means that there is a deformation-induced
potential difference between the interface and donor
sites, in spite of the homogeneous nature of the 
considered deformation component. 
The appearance of that effective potential difference
is due to two factors: the nontrivial valley structure
of the Herring-Vogt Hamiltonian, see Eq.~\eqref{eq:herringvogt},
and the different valley compositions of the two localized 
orbitals $\ket{\text{i}}$ and $\ket{\text{d}}$,
see Eq.~\eqref{eq:valleycompositions}.

Let us illustrate that claim, and 
the corresponding physical mechanism,
with a simple example. 
Take a longitudinal phonon propagating along the x axis. 
This case corresponds to $\epsilon_{yy} = \epsilon_{zz} = 0$ 
and a finite $\epsilon_{xx}$.
Hence, according to Eq.~\eqref{eq:herringvogt},
the conduction-band edges in the $x$ and $\bar{x}$ valleys
are raised by the uniaxial deformation potential 
$\Xi_u \epsilon_{xx}$, whereas the conduction band edges
in the other four valleys are not affected. 
Then, this effective potential in the $x$ and $\bar{x}$ valleys
is felt differently by $\ket{\text{i}}$ and $\ket{\text{d}}$: 
the state $\ket{\text{i}}$ has no weight in the $x$ and $\bar{x}$ valleys
[see Eq.~\eqref{eq:valleyi}],
therefore it does not feel the presence of the deformation;
the state $\ket{\text{d}}$,
however, has a total weight of $1/3$ in the $x$ and $\bar{x}$
valleys together [see Eq.~\eqref{eq:valleyd}], 
and hence the deformation raises its potential 
energy by $\Xi_u \epsilon_{xx} / 3$.
Therefore we conclude that a homogeneous deformation 
indeed induces a potential energy difference between
the interface orbital and the donor orbital. 
Furthermore, our 
argument translates to an effective electron-phonon
coupling Hamiltonian with a nontrivial part of
$-\Xi_u \epsilon_{xx} \sigma_\text{z}/6$,
in line with the corresponding term in Eq.~\eqref{eq:ephc}.

\subsection{Results}

To obtain the orbital relaxation time, 
Fermi's Golden Rule \eqref{eq:fgr} is  evaluated as
\bean
\frac{1}{T_\text{1,o}} &=& 
\frac{\epsilon_\text{o} V_t^2 \,  \Xi_u^2 }{60 \pi \hbar^4 \rho}
\left(
\frac 2 {3 v_\text{L}^5} + \frac{1}{v_\textrm{T}^5}
\right),
\label{eq:t1cresult}
\eean
where we used 
\bean
\braket{\text{g} | \sigma_z | \text{e}} = V_t /\epsilon_\text{o}.
\eean

At the ionization point, where $\epsilon_\text{o} = V_t$, 
and using the working-point parameters specified 
in Table \ref{tab:params},
the orbital relaxation rate is estimated as 
$1/T_\text{1,o} \approx 0.49\,\text{MHz}$,
corresponding to a relaxation time of $T_\text{1,o} \approx 2.1 \, \mu$s.

Upon detuning from the ionization point, 
the charge qubit energy splitting $\epsilon_\text{o}$ increases, 
and therefore, according to Eq.~\eqref{eq:t1cresult},
the relaxation speeds up.
This is interpreted as the result of a competition between
three effects. 

First, relaxation should slow down 
upon detuning from the ionization point
because the charge qubit basis states
$\ket{\text{g}}$ and $\ket{\text{e}}$ become more localized,
which suppresses the 
relevant matrix element $\braket{\text{g} | \sigma_z | \text{e}}$.
Second, relaxation should be enhanced upon detuning from the 
ionization point, as the charge qubit energy splitting 
$\epsilon_\text{o}$ increases, 
and therefore the density of states of the available phonons also 
increases. 
These two mechanisms exactly cancel each other. 

The fact that the relaxation speeds up upon detuning from
the ionization point is therefore a consequence of a third fact:
the vacuum fluctuation of the strain of a phonon mode
with energy $\epsilon_\text{o}$ is proportional to 
$\sqrt{\epsilon_\text{o}}$;
that follows from Eqs.~\eqref{eq:strainunexpanded} and \eqref{eq:homogeneousdeformation}, and the
energy conservation 
condition 
$\epsilon_\text{o} = \hbar v_\lambda q$
embedded in Fermi's Golden Rule \eqref{eq:fgr}. 
The quadratic form of Fermi's Golden Rule then implies 
a $1/T_\text{1,o} \propto \epsilon_\text{o}$ dependence
due to this factor, which does indeed appear in 
our result \eqref{eq:t1cresult}.

\subsection{Valley-enhanced relaxation}
\label{sec:valleyenhanced}

We wish to highlight the fact that the nontrivial features
of the setup associated to the valley degree of freedom 
boost the orbital relaxation process, 
and will play the same role in the flip-flop relaxation
and electron spin relaxation processes to be described below. 
In that sense, all these can be considered \emph{valley-enhanced}
relaxation processes. 
Our argument supporting that claim is as follows. 
The two relevant features are
(i) 
the nontrivial valley structure of the electron-phonon interaction,
and
(ii) the different valley compositions of the 
localized charge states $\ket{\text{i}}$ and $\ket{\text{d}}$. 
In the absence of any of these two ingredients, 
the first, homogeneous-deformation term in the plane-wave expansion
$e^{i\vec q \cdot \vec r} \approx 1 + i \vec q \cdot \vec r + \dots$ of
the strain tensor \eqref{eq:strainunexpanded} 
would give a vanishing contribution
to the relaxation rate, 
and therefore the relaxation rate would
be suppressed by a factor of $(q d)^2$.
Using the parameter values of Table \ref{tab:params},
that 
factor has the value of $(q d)^2 \approx 3 \times 10^{-3}$
[$(q d)^2 \approx  10^{-2}$]
for longitudinal [transverse] phonons.

In conclusion, in this section we have described 
a phonon-emission-mediated
orbital relaxation process,
characteristic of a charge qubit formed by 
a gate-tuned electron located between its donor atom 
and a nearby interface.
In particular, we have shown that the relaxation process is
enhanced by
(i)  the nontrivial valley structure of  
the electron-phonon interaction and 
(ii) 
the different valley compositions of the two 
orbital wave functions forming the charge qubit.

\section{Flip-flop relaxation}
\label{sec:ffq}

\begin{figure*}
\begin{center}
\includegraphics[width=2\columnwidth]{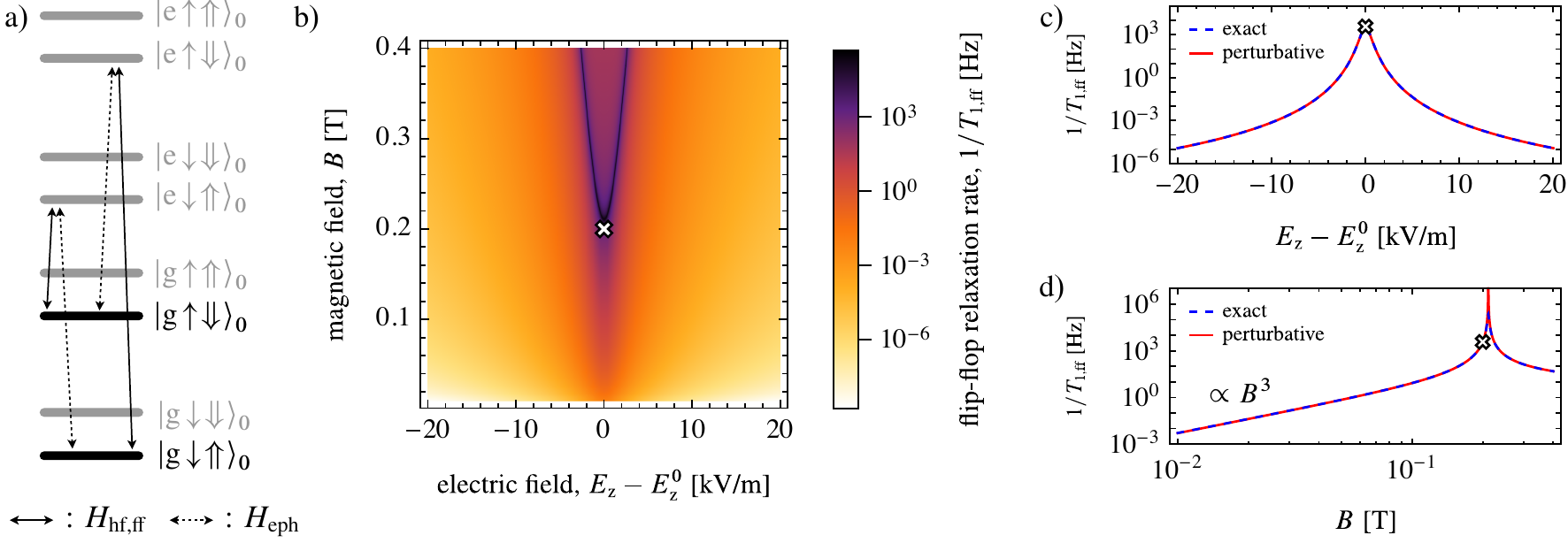}
\end{center}
\caption{(Color online) Flip-flop relaxation via
spontaneous phonon emission. 
(a) Hyperfine flip-flop matrix elements (solid arrows)
and electron-phonon matrix elements (dashed arrows)
enabling flip-flop relaxation. 
(b) Flip-flop relaxation rate $1/T_\text{1,ff}$
as a function of electric and magnetic fields. 
The white cross corresponds to the working point 
in Table \ref{tab:params};
further parameter values are also specified there.
See text for definition of the flip-flop qubit basis states.
(c,d) Blue dashed: horizontal/vertical cut of (b) through
the working point (white cross in b,c,d).
Red solid: analytical perturbative result, Eq.~\eqref{eq:gamma1ffpert}.
}
\label{fig:gamma1ff}
\end{figure*}

Here, we use the model described in sections \ref{sec:ffqintro}
and \ref{sec:charge}
to characterize 
the phonon-emission-mediated relaxation process
from the flip-flop qubit 
excited state $\ket{\text{g}\!\downarrow \Uparrow}$
to its ground state $\ket{\text{g}\!\uparrow \Downarrow}$.
This process is labelled in Fig.~\ref{fig:setup}c as  $T_\text{1,ff}$.
The characteristic time scale of this process for an isolated 
P donor at low temperature 
in bulk silicon has been estimated\cite{Pines} as 
$T_\text{1,ff} \sim 10^4$ s.
Here we show that this time scale can decrease by 
approximately 8 orders of magnitudes,
that is, $T_\text{1,ff} \sim 100 \, \mu$s is
possible,
if the flip-flop qubit is tuned to couple strongly to 
electric fields.

The flip-flop relaxation mechanism is 
visualized using the level diagram in Fig.~\ref{fig:gamma1ff}a.
It can be thought of as a two-step or second-order process,
in which matrix elements of the flip-flop
part of the hyperfine interaction $H_\text{hf,ff}$,
depicted as solid arrows in Fig.~\ref{fig:gamma1ff}a,
and matrix elements of 
the electron-phonon interaction $H_\text{eph}$,
denoted as dashed arrows in Fig.~\ref{fig:gamma1ff}a,
provide relaxation paths via virtual intermediate states.

Our calculation of $T_\text{1,ff}$ follows the preliminaries
and derivation steps of the calculation of 
$T_\text{1,o}$ in the previous
section. 
For the flip-flop relaxation rate, Fermi's Golden Rule 
reads
\begin{align}
\label{eq:ff_fgr}
\frac 1 {T_\text{1,ff}} = 
\frac{2\pi}{\hbar} 
\sum_{\vec q,\lambda}
\left| 
\braket{\text{g}\!\downarrow \Uparrow, \vec q \lambda 
| H_\text{eph,o} | \text{g}\!\uparrow \Downarrow, 0}
\right|^2 
\delta(\epsilon_\text{ff} - \hbar v_\lambda q).
\end{align}
As long as $H_\text{hf,ff}$ is a perturbation
of $H_\text{sec}$, we can use first-order perturbation
theory to obtain analytical approximate expressions 
for the qubit basis states
$\ket{\text{g}\!\downarrow \Uparrow}$ and
$\ket{\text{g}\!\uparrow \Downarrow}$ in
terms of the 8 unperturbed energy eigenstates.
In fact, the form of  $H_\text{hf,ff}$ guarantees that 
the flip-flop qubit basis
states are linear combinations of the 
4 unperturbed energy eigenstates
$\ket{\text{g}\!\downarrow \Uparrow}_0$,
$\ket{\text{g}\!\uparrow \Downarrow}_0$,
$\ket{\text{e}\!\downarrow \Uparrow}_0$, and
$\ket{\text{e}\!\uparrow \Downarrow}_0$.
The flip-flop relaxation rate is then readily evaluated from
Eq.~\eqref{eq:ff_fgr}  as:
\bean
\label{eq:gamma1ffpert}
\frac{1}{T_\text{1,ff}} = 
\frac{A^2\Xi_u^2V_\text{t}^4\epsilon_B^3 }
{240 \pi \hbar^4 \rho \epsilon_\text{o}^2 \left(\epsilon_\text{o}^2-\epsilon_B^2\right)^2}
\left(
\frac 2 {3 v_\text{L}^5} + \frac{1}{v_\textrm{T}^5}
\right).
\eean
Here, 
$\epsilon_B=h \gamma_\text{e} B$,
and Eq.~\eqref{eq:gamma1ffpert} shows the leading-order 
result in the small parameters 
$A/\left(\epsilon_\text{o} - \epsilon_B\right),\,
h \gamma_\text{n} B / \left(\epsilon_\text{o} - \epsilon_B\right)
\ll 1$.

This result can also be expressed in terms of the 
orbital relaxation time:
\bean
\frac{1}{T_\text{1,ff}} = \frac 1 4
\frac{A^2 V_\text{t}^2 \epsilon_B^3}
{\epsilon_\text{o}^3
\left(
	\epsilon_\text{o}^2 - \epsilon_B^2
\right)^2}
\frac{1}{T_\text{1,o}},
\eean
taking a particularly simple approximate form 
in the vicinity of the proposed working point,
where 
the electron is placed halfway between the interface and the donor
and the energy splittings of 
the charge qubit and flip-flop qubit
are similar 
($V_\text{t} \approx 
\epsilon_\text{o} \approx \epsilon_B$):
\bean
\frac{1}{T_\text{1,ff}} \approx
\left(\frac{A/4}
{
	\epsilon_\text{o} - \epsilon_B
}
\right)^2
\frac{1}{T_\text{1,o}}.
\label{eq:gamma1ffsimple}
\eean
Note that this result corresponds to the special case when 
the leftmost virtual transition of Fig.~\ref{fig:gamma1ff}a
dominates the relaxation process.

With the parameter values in 
Table \ref{tab:params}, 
from Eq.~\eqref{eq:gamma1ffpert} we obtain $1/T_\text{1,ff} \approx 3.7 \, \text{kHz}$,
implying a flip-flop relaxation time of 
$T_\text{1,ff} \approx 270 \, \mu\text{s}$.
This value is approximately 
8 orders of magnitude shorter than the $10^4$ s
time scale that was estimated for an
on-donor electron in bulk by Ref.~\onlinecite{Pines}.
The reason for the fast relaxation at the 
proposed working point of the flip-flop qubit is
is twofold. 
First, the working point is chosen with the goal of 
optimizing the speed of electrically driven qubit transitions:
$\epsilon_\textrm{o} - \epsilon_B$, 
which appears as an energy denominator in the
perturbative description of the leftmost virtual process
of Fig.~\ref{fig:gamma1ff}, is chosen to be
relatively small ($\sim h \times 300$ MHz), 
so that the qubit excited state 
$\ket{\text{g}\!\uparrow \Downarrow}$ 
has a relatively large, hyperfine-mediated 
admixture with the unperturbed energy eigenstate
$\ket{\text{e}\!\downarrow \Uparrow}_0$.
Second, this flip-flop relaxation process is valley-enhanced,
in a similar sense as described in section \ref{sec:valleyenhanced}.
That is, due to the nontrivial valley structure of the 
electron-phonon Hamiltonian and the different valley compositions
of the involved electronic orbitals $\ket{\text{i}}$ and $\ket{\text{d}}$, 
even a uniform phonon-induced deformation is capable 
to induce relaxation.

If the charge-qubit splitting $\epsilon_\text{o}$
is much larger than the electronic Zeeman splitting $\epsilon_B$, 
then 
Eq.~\eqref{eq:gamma1ffpert}
implies the power-law relation
$1/T_\text{1,ff} \propto B^3$,
see also Fig.~\ref{fig:gamma1ff}d. 
The 3rd power arises as a sum $1+2$,
where the terms, respectively, are associated to 
the strain vacuum fluctuations and 
the density of states of three-dimensional acoustic phonons.
This is analogous to the low-temperature limiting case
$1/T_\text{1,ff} \propto B^3$
of the relaxation mechanism considered in Ref.~\onlinecite{Pines}
for on-donor electrons in bulk:
even though the mechanisms considered here and there
are different, in both cases a homogeneous deformation is
responsible for the relaxation.

In Fig.~\ref{fig:gamma1ff}b, we show the dependence of 
the qubit relaxation rate $1/T_\text{1,ff}$
on the gate-induced electric field 
and the magnetic field.
To obtain this result, we first numerically computed the 
eigenvalues and eigenvectors of $H$.
Then, we identified the flip-flop qubit ground (excited) state
as the energy eigenstate having the largest
overlap with $\ket{\text{g}\!\downarrow \Uparrow}_0$
($\ket{\text{g}\!\uparrow \Downarrow}_0$).
Finally, we evaluated the relaxation rate
according to Eq.~\eqref{eq:ff_fgr}.

The key features in Fig.~\ref{fig:gamma1ff}b
are as follows.
(i) The qubit relaxation rate is strongly suppressed at low 
magnetic fields, 
due to the above-discussed $1/T_\text{1,ff} \propto B^3$
dependence.
(ii) The qubit relaxation rate is maximal, 
taking values around 1 MHz, along the 
upward-bending hyperbola, which corresponds to 
$\epsilon_\text{o} \approx \epsilon_B$, and,
therefore, nonperturbative mixing of $\ket{\text{g}\!\uparrow \Downarrow}$
and $\ket{\text{e}\!\downarrow \Uparrow}$.
Hence this relaxation rate of 1 MHz reflects 
the orbital relaxation rate. 

Comparison of the numerical results of
Fig.~\ref{fig:gamma1ff}b and the perturbative, analytical 
expression \eqref{eq:gamma1ffpert} is shown
in Fig.~\ref{fig:gamma1ff}c,d.
In Fig.~\ref{fig:gamma1ff}c, the  dashed blue line 
shows a horizontal cut of Fig.~\ref{fig:gamma1ff}b
through the working point (white cross),
whereas the solid red line is the 
analytical result.
A similar comparison, corresponding to a vertical cut
of Fig.~\ref{fig:gamma1ff}b through the working point,
is shown in Fig.~\ref{fig:gamma1ff}d. 
Note that 
in Fig.~\ref{fig:gamma1ff}d ,
for magnetic fields slightly higher than the 
working-point magnetic field,
the analytical result 
deviates from the numerical one and diverges;
that behavior is an artefact arising from 
the breakdown of first-order perturbation theory.

In conclusion, we have proposed a valley-enhanced
relaxation mechanism of the flip-flop qubit,
calculated its characteristic relaxation time $T_\text{1,ff}$, 
and found a relatively short, $\sim 100\, \mu$s
time scale in the proposed working point.
This is partly due to the presence of a low-lying 
orbital that is utilized to enhance the coupling of the 
qubit to the electric field. 
Another factor boosting the relaxation process is the 
absence of dipole suppression (see section \ref{sec:valleyenhanced}):  
thanks to the nontrivial valley
structure of the electron-phonon interaction and the 
involved electronic orbitals $\ket{\text{i}}$ and $\ket{\text{d}}$, 
a homogeneous deformation can induce an effective
potential difference between the two orbitals, and hence
lead to efficient relaxation.

\section{Electron spin relaxation}
\label{sec:spin}

A further  process, leading to leakage from 
the flip-flop qubit subspace, is electron spin relaxation
(henceforth \emph{spin relaxation}, for short):
this  is shown in Fig.~\ref{fig:setup}c, labelled 
as $T_\text{1,s}$.
We first describe a valley-enhanced spin-relaxation
mechanism that is enabled by spin-orbit interaction; 
more precisely, by spin-dependent
electron  tunnelling between the two 
localized orbitals $\ket{\text{i}}$ and
$\ket{\text{d}}$. 
We also discuss an alternative valley-enhanced relaxation  
mechanism, which is enabled by the feature that
the $g$-tensors characterizing the localized
orbitals
$\ket{\text{i}}$ and $\ket{\text{d}}$ are, in general, different and 
anisotropic\cite{Rahman_gfactor}.

\subsection{Spin relaxation due to spin-dependent 
tunneling}
\label{subsec:spinorbit}

First, we incorporate spin-orbit interaction to our 
$8\times 8$ model Hamiltonian described in section \ref{sec:ffqintro}.
For simplicity, we assume that the setup 
is cylindrically symmetric around the z axis. 
We claim that this symmetry condition, together with the 
condition that the spin-orbit Hamiltonian must be 
invariant under time reversal, 
imply the following simple form for the spin-orbit Hamiltonian:
\bean
\label{eq:soi}
H_\text{so} = V_\text{s} \sigma_\text{y} S_\text{z},
\eean
where $V_\text{s}$ is real. 
Naturally, this Hamiltonian excludes the nuclear-spin operators. 
Furthermore, since $\sigma_\text{y}$ is 
an off-diagonal matrix, $H_\text{so}$ 
describes spin-dependent tunneling between 
the two orbitals $\ket{\text{i}}$ and $\ket{\text{d}}$.

The proof of Eq.~\eqref{eq:soi}, inspired by a related argument
of Ref.~\onlinecite{Danon_spinblockade},
is as follows. 
In principle, the spin-orbit Hamiltonian can be expanded
in terms of products of charge-qubit Pauli matrices including
the unit matrix $\sigma_0$, 
and the three spin Pauli matrices:
$H_\text{so} = 
\sum_{i=0,\text{x},\text{y},\text{z}} 
\sum_{j=\text{x},\text{y},\text{z}} 
\alpha_{ij} \sigma_i S_j$, 
where $\alpha_{ij}$ represent 12 unknown coefficients.
Then, the condition of time reversal invariance renders 
9 of the coefficients zero, 
$\alpha_{0j} = \alpha_{\text{x} j} = \alpha_{\text{z} j} = 0$,
for the following reason. 
Time reversal is represented as $T = 2 i S_\text{y} K$
with $K$ being the complex conjugation, 
therefore 
the spin matrices $S_\text{x}$, $S_\text{y}$, and $S_\text{z}$
change sign under time reversal, whereas
the real matrices 
$\sigma_0$, $\sigma_\text{x}$ and $\sigma_\text{z}$ keep their signs.
That implies that the only 
charge-qubit Pauli matrix allowed in the
spin-orbit Hamiltonian is $\sigma_\text{y}$:
$H_\text{so} = 
\sigma_\text{y} \left(
\alpha_{\text{y} \text{x}} S_\text{x}
+
\alpha_{\text{y} \text{y}} S_\text{y}
+
\alpha_{\text{y} \text{z}} S_\text{z}
\right)
$.
However, a finite value of either $\alpha_\text{yx}$ or
$\alpha_\text{yy}$ would specify a certain direction in the xy plane,
which is disallowed by the cylindrical symmetry of the setup
around the z axis. 
With the identification $V_\text{s} \equiv \alpha_\text{yz}$,
this concludes the proof of Eq.~\eqref{eq:soi}.
A quantitative characterization of $V_\text{s}$ could be obtained from
microscopic, e.g., tight-binding\cite{Rahman_gfactor}, 
simulations, incorporating the
nanostructure geometry and 
spin-orbit interaction.
Here, we treat $V_\text{s}$ as a phenomenological parameter.

\begin{figure*}
\begin{center}
\includegraphics[width=2\columnwidth]{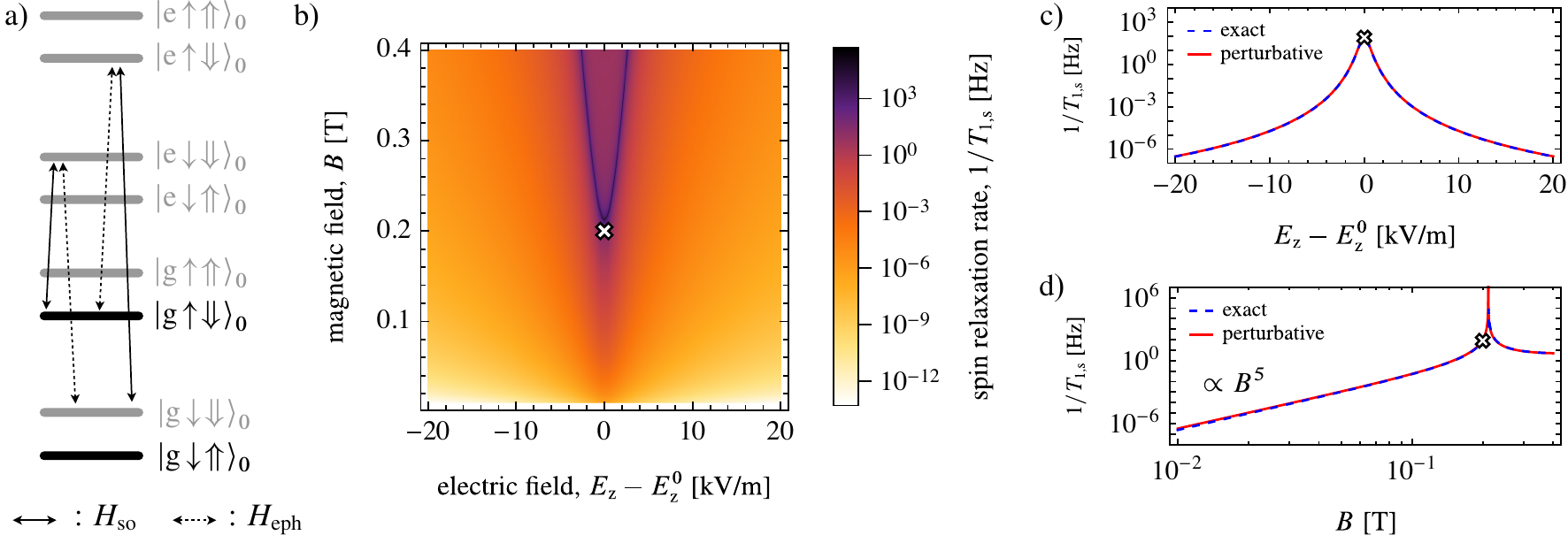}
\end{center}
\caption{(Color online) 
Leakage from the flip-flop qubit subspace: 
electron spin relaxation due to spin-dependent tunneling and 
spontaneous phonon emission.
(a) Spin-dependent tunneling matrix elements (solid arrows)
and electron-phonon matrix elements (dashed arrows)
enabling electron-spin relaxation. 
(b) Electron spin relaxation rate $1/T_\text{1,s}$
as a function of electric and magnetic fields,
for $V_\text{s}/h = 10$ MHz. 
The white cross corresponds to the working point 
in Table \ref{tab:params};
further parameter values are also specified there.
(c,d) Blue dashed: horizontal/vertical cut of (b) through
the working point (white cross).
Red solid: analytical perturbative result, Eq.~\eqref{eq:gamma1spert}.
}
\label{fig:gamma1s}
\end{figure*}

Having the spin-orbit Hamiltonian at hand, 
we now propose the spin relaxation mechanism 
it enables.
The mechanism is analogous to the flip-flop relaxation, 
and is visualized using the level diagram in Fig.~\ref{fig:gamma1s}a. 
Here, we parametrize the magnetic-field orientation
via its polar angle $\theta$: $\vec B = B(\sin \theta, 0 , \cos\theta)$,
but disregard any orbital effects caused by $\vec B$.
Furthermore, recall that the arrows 
in our state notation (for example, in $\ket{\text{g}\!\uparrow \Downarrow}_0$)
correspond to spin alignments
with respect to the external magnetic field, not with respect to z. 
Then, we conclude that $H_\text{so}$ mixes
 the unperturbed state $\ket{\text{g}\!\uparrow \Downarrow}_0$
 with $\ket{\text{e}\!\downarrow \Downarrow}_0$.
 This mixing is depicted as the left solid
 arrow in Fig.~\ref{fig:gamma1s}a. 
Since, in turn, $\ket{\text{e}\!\downarrow \Downarrow}_0$ 
is connected to 
 $\ket{\text{g}\!\downarrow \Downarrow}_0$
 by the electron-phonon interaction
 (left dashed arrow in Fig.~\ref{fig:gamma1s}a), 
 we conclude that the spin-orbit interaction 
does indeed enable spin relaxation. 
A similar two-step process contributing to 
spin relaxation is depicted by the right
solid and dashed arrows. 

The spin relaxation rate arising from these
second-order processes can be calculated as
\begin{align}
\label{eq:spin_fgr}
\frac 1 {T_\text{1,s}} = 
\frac{2\pi}{\hbar} 
\sum_{\vec q,\lambda}
\left| 
\braket{\text{g}\!\downarrow \Downarrow, \vec q \lambda 
| H_\text{eph,o} | \text{g}\!\uparrow \Downarrow, 0}
\right|^2 
\delta(\epsilon_\text{s} - \hbar v_\lambda q).
\end{align}
where the states $\ket{\text{g}\!\downarrow \Downarrow}$ 
and $\ket{\text{g}\!\uparrow \Downarrow}$ are perturbed by 
the spin-orbit interaction,
in analogy to Eq.~\eqref{eq:ff_fgr},
where the states are perturbed by the flip-flop terms of the
hyperfine interaction.
Furthermore, $\epsilon_s$ is the energy splitting
between the energy eigenstates
$\ket{\text{g}\!\uparrow \Downarrow}$ and
$\ket{\text{g}\!\downarrow \Downarrow}$.
 
Using first-order perturbation theory to account for 
the spin-orbit-induced mixing of the unperturbed states,
we find
\bean
\label{eq:gamma1spert}
\frac{1}{T_\text{1,s}} = 
\frac{V_\text{s}^2\sin^2\theta\,\Xi_u^2V_\text{t}^2\epsilon_B^5 }
{15 \pi \hbar^4 \rho \epsilon_\text{o}^2 \left(\epsilon_\text{o}^2-\epsilon_B^2\right)^2}
\left(
\frac 2 {3 v_\text{L}^5} + \frac{1}{v_\textrm{T}^5}
\right),
\eean
which shows the leading-order 
result in the small parameters 
$A/\left(\epsilon_\text{o} - \epsilon_B\right)
,\,
h \gamma_\text{n} B /\left(\epsilon_\text{o} - \epsilon_B\right)
,\,
V_\text{s} / \left(\epsilon_\text{o} - \epsilon_B\right)
\ll 1$.

 Expressed with the orbital relaxation rate:
 \bean
 \frac{1}{T_\text{1,s}} = 
 \frac{4V_\text{s}^2\sin^2\theta\, \epsilon_B^5}
{
\epsilon_\text{o}^3
\left(
	\epsilon_\text{o}^2 - \epsilon_B^2
\right)^2}
 \frac{1}{T_\text{1,o}}
 \eean
 In the vicinity of the proposed working point where
 $\epsilon_\text{o} \approx \epsilon_B$, 
 this is approximated
 as
\bean
\frac{1}{T_\text{1,s}} \approx
\left(\frac{\frac 1 2 V_\text{s}\sin\theta}
{
	\epsilon_\text{o} - \epsilon_B
}
\right)^2
\frac{1}{T_\text{1,o}}.
\eean

At weak magnetic fields, $\epsilon_B \ll \epsilon_\text{o}$, 
the spin-relaxation rate in Eq.~\eqref{eq:gamma1spert} 
follows the power-law relation
$1/T_\text{1,s} \propto B^5$,
see Fig.~\ref{fig:gamma1s}d;
this is a stronger dependence then the $1/T_\text{1,ff} \propto B^3$
seen in the previous section, 
and the difference is due to van Vleck 
cancellation\cite{vanVleck-cancellation,Khaetskii_2001}. 

In Fig.~\ref{fig:gamma1s}b, we show the dependence of 
the spin relaxation rate $1/T_\text{1,s}$
on the gate-induced electric field 
and the magnetic field, in analogy
to Fig.~\ref{fig:gamma1ff}, 
using
the numerically computed
eigenvalues and eigenvectors of $H+H_\text{so}$. 
To produce this plot, the magnetic field is assumed to be aligned with the x axis ($\theta = \pi/2$),
and for not having a calculated or measured
value for the spin-dependent tunneling energy $V_\text{s}$, 
we used an arbitrary value $V_\text{s}/h = 10$ MHz. That implies that even though the parameter dependencies
of the spin relaxation rate shown in 
Fig.~\ref{fig:gamma1s}b,c,d are expected to be 
accurate, the actual numerical values should not be 
regarded as predictions.
Having a realistic estimate $V_\text{s,est}$ 
for the spin-dependent tunneling amplitude, 
the plotted results could be rescaled 
to provide numerical 
predictions by multiplying with $(V_\text{s,est}/10 \, \text{MHz})^2$.

The key features in Fig.~\ref{fig:gamma1s}b
are analogous to those of the flip-flop relaxation. 
(i) The spin relaxation rate is strongly suppressed at low 
magnetic fields, 
due to the above-discussed $1/T_\text{1,s} \propto B^5$
dependence.
(ii) The spin relaxation rate is  maximal
along an 
upward-bending hyperbola, corresponding to 
$\epsilon_\text{o} \approx \epsilon_B$, and,
therefore, nonperturbative mixing of $\ket{\text{g}\!\uparrow \Downarrow}$
and $\ket{\text{e}\!\downarrow \Downarrow}$.
(iii) 
For the working point specified in Table \ref{tab:params}
(white cross in Figs. \ref{fig:gamma1s}b,c,d),
the numerical value for the spin relaxation rate
evaluated from Eq.~\eqref{eq:gamma1spert} is 
$1/T_{1,\text{s}} \approx 98\,\text{Hz}$,
that is, the spin relaxation time is
$T_{1,\text{s}} \approx 10 \, \text{ms}$.

A comparison between the exact and perturbative 
results, for a horizontal (vertical) cut of Fig.~\ref{fig:gamma1s}b
across the working point, is shown in 
Figs.~\ref{fig:gamma1s} c (d).

\subsection{Spin relaxation due to $g$-tensor modulation}

We conclude the list of valley-enhanced relaxation mechanisms
with spin relaxation due to 
$g$-tensor modulation. 
This process is allowed if
the $g$-tensors associated to the 
localized orbitals $\ket{\text i}$ and $\ket{\text d}$
are different, and, e.g., that of $\ket{\text i}$ 
is anisotropic\cite{Rahman_gfactor,Tosi}.
In that case, a phonon, corresponding to an effective 
potential difference between the two localized orbitals,
redistributes the electron between the two locations and
thereby changes the $g$-tensor.
In general, that implies that both the length and the direction
of the effective Zeeman field felt by the electron 
changes, leading to spin relaxation.

We focus on the simple case when the $g$-tensors show
cylindrical symmetry along the growth direction z.
Importantly, in this case, this relaxation process can be 
avoided by a perfect in-plane or out-of-plane alignment of
the external magnetic field $\vec B$.
We further assume that $g$-tensor anisotropy is 
present only at the interface. 
That anisotropy is incorporated in our model as the 
perturbation term
\bean
H_{g\text{tm}} = 
\left(\frac{1+\sigma_\text{z}}{2}\right)
h\gamma_\text{e}
\vec B \begin{pmatrix}
  \Delta_\gamma^\perp & 0 & 0 \\
  0 & \Delta_\gamma^\perp & 0 \\
  0 & 0 & \Delta_\gamma^\parallel
 \end{pmatrix}
\vec S.
\eean
Tight-binding nanostructure models predict\cite{Rahman_gfactor}
that the typical absolute value of the
relative $g$-tensor anisotropy parameters
is in the range
$| \Delta_\gamma^\perp |, | \Delta_\gamma^\parallel |  \in 
[0.1,1]\%$.

To evaluate the corresponding spin relaxation time $T_{1,\text{s}}$, 
we follow the same procedure as in section \ref{subsec:spinorbit},
but now instead of $H_\text{so}$, 
we use $H_{g\text{tm}}$ as the perturbation in the Hamiltonian.
(Note that the two mechanisms do interfere in general; 
we disregard that here and discuss their effects separately
for simplicity.)
The leading-order perturbative result, expressed using
the orbital relaxation time $T_{1,\text{o}}$, reads
\bean
\label{eq:gamma1spertgtm}
 \frac{1}{T_\text{1,s}} = 
\frac{1}{16}
 \frac{[(\Delta_\gamma^\perp-\Delta_\gamma^\parallel)\sin (2\theta)]^2 V_\text{t}^2\epsilon_B^5}
{
\epsilon_\text{o}^3
\left(
	\epsilon_\text{o}^2 - \epsilon_B^2
\right)^2}
 \frac{1}{T_\text{1,o}}.
\eean
 In the vicinity of the proposed working point where
 $V_\text{t} \approx \epsilon_\text{o} \approx \epsilon_B$, 
 this is approximated
 as
\bean
\label{eq:gamma1spertgtm2}
\frac{1}{T_\text{1,s}} \approx
\left[\frac{\epsilon_B (\Delta_\gamma^\perp-\Delta_\gamma^\parallel)\sin (2\theta)/8}
{
	\epsilon_\text{o} - \epsilon_B
}
\right]^2
\frac{1}{T_\text{1,o}}.
\eean
With the parameters in Table \ref{tab:params},
we estimate the maximal spin relaxation rate, corresponding to 
a B-field polar angle of $\theta = \pi/4$, as
$1/T_\text{1,s} \approx 140\,\text{Hz}$,
implying a spin relaxation time of $T_\text{1,s} \approx 7 \, \text{ms}$.
Recall that under our presumptions, 
this mechanism can be fully suppressed by aligning
the magnetic field in the xy plane or along the z axis;
this appears explicitly in the results \eqref{eq:gamma1spertgtm}
and \eqref{eq:gamma1spertgtm2} via the factor $\sin(2\theta)$.

In conclusion, we proposed that spin-orbit-induced
spin-dependent tunneling between the localized
charge states $\ket{\text{i}}$ and $\ket{\text{d}}$ can
induce a valley-enhanced electron spin relaxation 
process, 
which leads to leakage from the flip-flop qubit subspace,
and we expressed the parameter dependence
of the corresponding relaxation rate. 
We also discussed how the different $g$-tensors characterizing
the two localized charge states $\ket{\text{i}}$ and $\ket{\text{d}}$ 
can contribute to electron spin relaxation.

\section{Discussion}

\subsection{Electrically driven spin resonance}

The spin relaxation process described in 
section \ref{subsec:spinorbit} is allowed 
by the spin-orbit-induced spin-dependent 
tunneling matrix element $V_\text{s}$.
The same matrix element could also be utilized 
for electrically driven electron spin resonance:
an ac voltage component on the top gate produces
an ac electric field $E_\text{ac}$ along z, 
which provides the same couplings as
the electron-phonon matrix elements 
depicted as dashed arrows in Fig.~\ref{fig:gamma1s}a, 
and thereby drives coherent transitions between
$\ket{\text{g}\!\downarrow \Downarrow}$ 
and
$\ket{\text{g}\!\uparrow \Downarrow}$.
The corresponding Rabi frequency reads
\bean
\label{eq:edsr}
f_\text{s,Rabi}=\frac{\epsilon_B V_\text{s}  V_\text{t} \sin \theta  \, e E_\text{ac} d}{2h \epsilon_\text{o}(\epsilon_\text{o}^2-\epsilon_B^2)}.
\eean
This result is obtained via the following steps:
we
(i) expressed the energy eigenstates
of the $8\times 8$ model Hamiltonian 
$H_\text{sec} + H_\text{hf,ff} + H_\text{so}$
using first-order perturbation
theory in $H_\text{hf,ff} + H_\text{so}$, 
(ii) projected the driving Hamiltonian 
$H_\text{ac} = e E_\text{ac} z \sin (\omega t)$
onto the two-dimensional subspace spanned by
the perturbed energy eigenstates 
$\ket{\text{g}\!\downarrow \Downarrow}$ 
and
$\ket{\text{g}\!\uparrow \Downarrow}$,
and
(iii) read off the Rabi frequency as the amplitude of the 
transverse driving term in the resulting two-dimensional
Hamiltonian. 
In the vicinity of the proposed working point, 
$V_\text{t} \approx \epsilon_\text{o} \approx \epsilon_B$, 
and in the presence of an in-plane magnetic field ($\theta = \pi/2$)
the result \eqref{eq:edsr} simplifies to 
\bean
f_\text{s,Rabi} = \frac{V_\text{s} 
e E_\text{ac} d}{4h (\epsilon_\text{o} - \epsilon_B)}.
\eean
Using the parameter values in Table \ref{tab:params},
and assuming an in-plane magnetic field ($\theta = \pi/2$),  
we find $f_\text{s,Rabi} \approx 0.89\,\text{MHz}$. 

For comparison, we provide the analogous
result for the Rabi frequency of
the electrically driven transitions of the flip-flop qubit:
\bean
\label{eq:ffrabi}
f_\text{ff,Rabi}=\frac{A V_\text{t}^2 e E_\text{ac} d}{4h\epsilon_\text{o}(\epsilon_\text{o}^2-\epsilon_B^2)}.
\eean
In the vicinity of the proposed working point, 
$V_\text{t} \approx \epsilon_\text{o} \approx
\epsilon_B$, the result \eqref{eq:ffrabi} simplifies to 
\bean
\label{eq:ffrabisimple}
f_\text{ff,Rabi} = \frac{A e E_\text{ac} d}{8h(\epsilon_\text{o} - \epsilon_B)},
\eean
and thereby we recover the corresponding 
result of Ref.~\onlinecite{Tosi}
[2 times the coupling rate in Eq.~(7) of Ref.~\onlinecite{Tosi}].

Finally, we highlight a potential use of 
electrically driven spin resonance in 
the nuclear-spin-based quantum processor proposed
in Ref.~\onlinecite{Tosi}. 
For that setup, a key ingredient is a magnetic drive of 
the donor electron spin via an ac magnetic field.
Creating such an ac magnetic field requires an 
extra element, for example, 
a microwave transmission line, in the setup.
Electrically driven spin resonance, 
allowed by a sufficiently strong
spin-dependent tunnel matrix element $V_\text{s}$
and driven by an ac gate voltage component, 
could substitute the ac magnetic field, and hence
reduce the complexity of the envisioned architecture. 
To assess the practical feasibility of electrically driven spin resonance,
a quantitative characterization of $V_\text{s}$ is required.

\subsection{Breaking of the approximate cylindrical symmetry 
affects spin relaxation}

In Ref.~\onlinecite{Tosi}, it is proposed that the interface-donor
tunneling amplitude $V_\text{t}$ is tuned
to the desired value by 
moving the interface orbital $\ket{\text{i}}$ away from 
the donor orbital $\ket{\text{d}}$ 
along the interface, using an appropriately
designed gate stack. 
Of course, in that case 
the approximate cylindrical symmetry assumed in 
our considerations of spin relaxation 
(section \ref{sec:spin}) is broken, and therefore 
our symmetry-based results have to be refined accordingly.

\subsection{Relaxation rates at finite temperature}

Here, we evaluated  relaxation
rates corresponding to zero temperature and
spontaneous phonon emission. 
Induced-emission and 
absorption rates at finite temperature $T>0$ 
are obtained by multiplying the
corresponding spontaneous-emission rates with 
the Bose-Einstein factor 
$n(\epsilon,T) = 1/(e^{\epsilon/k_\text{B} T}-1)$,
where $\epsilon$ is the energy of the involved phonons
and $k_\text{B}$ is the Boltzmann constant.

\subsection{Orbital relaxation: comparison to experiment}

A recent experiment\cite{Urdampilleta} reports 
an orbital relaxation time $T_\text{1,o} \approx 0.1 \, \mu\text{s}$ of
a charge qubit, formed in an effective double quantum dot system
in a silicon nanowire transistor, 
where one of the dots is presumably a single P donor,
while the other one is gate-defined. 
The quoted orbital relaxation time, measured at the 
charge-qubit anticrossing
point 
at a nominal charge-qubit energy splitting of 
$\epsilon_\text{o} = V_\text{t} =h \times 5.5$ GHz, 
can be compared to the corresponding prediction of our 
Eq.~\eqref{eq:t1cresult}, that is $T_\text{1,o} \approx 2.5 \, \mu\text{s}$.

Note that even though the two setups, studied in 
Ref.~\onlinecite{Urdampilleta}
and in this work, share their hybrid dot-donor character, there
are also important differences between them:
(i) The charge qubit in the experiment is formed by two electrons, 
in the (1,1)--(0,2) charge configuration, where the first (second)
integer is the number of electrons in the quantum dot (on the donor);
our result \eqref{eq:t1cresult} 
corresponds to the single-electron case. 
(ii) In the experiment, the quantum dot is formed at the 
corner of a nanowire, which presumably implies that the
valley composition of the occupied 
electronic state is different from that
described by Eq.~\eqref{eq:valleyi}, the latter corresponding to 
an electronic state at a (001) silicon/barrier interface.
(iii) In the experiment, the inhomogeneous dephasing time
of the charge qubit is comparable to its splitting at the 
(1,1)--(0,2) anticrossing. 
This indicates the presence of relatively strong electrical
noise affecting the charge qubit detuning or tunnel coupling.
Therefore, the measured orbital relaxation time 
should probably 
be understood as an average over a random ensemble of the
charge-qubit parameters. 

Note also that the orbital relaxation time $T_\text{1,o}$ 
was measured for a single setting of the charge qubit. 
Measuring $T_\text{1,o}$ as a function
of the charge-qubit parameters would allow
 a qualitative comparison with theoretically predicted trends, 
e.g., Eq.~\eqref{eq:t1cresult} of this work, and thereby 
help identifying the underlying relaxation mechanism.

\subsection{Prolonging the relaxation times}

(1) \emph{Controlling the 
valley composition of the donor orbital $\ket{\text{d}}$.}
In the valley-enhanced relaxation mechanisms described in 
this work, a key ingredient is the substantially different valley 
structure of the electronic wave functions
at the interface and donor sites, see Eq.~\eqref{eq:valleycompositions}.
Making the valley composition [Eq.~\eqref{eq:valleyd}] of the donor 
orbital 
more similar to that [Eq.~\eqref{eq:valleyi}] 
of the interface orbital would prolong 
the relaxation times. 
The even valley composition of $\ket{\text{d}}$ 
in Eq.~\eqref{eq:valleyd} 
might be altered by a number of mechanisms: 
for example, 
by static strain due to a finite germanium
concentration in the heterostructure\cite{Vrijen_PRA00,Tahan2002,Koiller_PRB02},
by the close vicinity of an interface\cite{Baena,Salfi_VlyInt_NM14},
or by an electric field\cite{Friesen-starkeffect}.
For example, placing the donor closer to an interface,
while keeping all other relevant parameters unchanged, would
bring the valley compositions of $\ket{\text{i}}$ and $\ket{\text{d}}$
closer to each other, and therefore
presumably prolong the relaxation times
considered here.
That speculation is supported by, e.g., the estimate 
in Ref.~\onlinecite{Salfi_VlyInt_NM14}, 
claiming that 
the z-valley population of a donor electron at $3.4$ nm 
below a silicon surface is  $\approx$40\%, 
in contrast to the bulk value 33\%.

(2) \emph{Optimizing the working point
via weakening the qubit-field interaction.}
In the vicinity of the working point 
of Table \ref{tab:params},
the estimated\cite{Tosi} time required for a cavity-mediated
$\sqrt{\text{SWAP}}$ two-qubit
gate is $\tau_{\sqrt{\text{SWAP}}} \approx 0.4 \, \mu\text{s}$.
This implies that 
the number of such operations 
performed during the flip-flop relaxation time 
is $T_\text{1,ff} / \tau_{\sqrt{\text{SWAP}}} \approx 680$.
In principle, this quality factor can be improved via, 
e.g., increasing the tunneling amplitude $V_\text{t}$,  
thereby weakening the hyperfine-induced
hybridization of $\ket{\text{g}\!\uparrow \Downarrow}_0$ with
$\ket{\text{e}\!\downarrow \Uparrow}_0$ (see Fig.~\ref{fig:gamma1ff}a),
and hence weakening the interaction between the flip-flop
qubit and the electric fields. 
For example, approximately a 
factor of 2 improvement of the above quality factor
can be achieved by the 
following adjustments. 
(i) The tunnel matrix element is reset to $V_\text{t} = 6.2$ GHz.
Essentially, this doubles the 
energy denominator $\epsilon_\text{o} - \epsilon_B$ 
in the 
flip-flop relaxation rate \eqref{eq:gamma1ffsimple}
as well as in the vacuum Rabi frequency;
the latter is obtained from \eqref{eq:ffrabisimple} by identifying
$E_\text{ac}$ with the cavity vacuum field.
As a result, $T_\text{1,ff}$ increases by
a factor of 4.
(ii) The magnetic field is reset such that 
the qubit-cavity detuning is halved.
As a result of (i) and (ii),  $\tau_{\sqrt{\text{SWAP}}}$ 
increases only with a factor of 2, without a significant change in 
the gate fidelity;
hence the quality factor $T_\text{1,ff}/\tau_{\sqrt{\text{SWAP}}}$ 
indeed doubles. 
In practice, an important consideration that should be added to the 
above procedure is the change of the inhomogeneous dephasing
time $T_\text{2,ff}^*$ of the flip-flop qubit with the adjustments,
with the goal of exploiting the expected long $T_\text{2,ff}^*$
times offered by the second-order clock transitions\cite{Tosi}.
In general, this necessitates a more complex optimization 
procedure.

\begingroup
\squeezetable
\begin{table}
\caption{\label{tab:timescales}
Time scales at the working point of Table \ref{tab:params}. Remarks: (1) Having no estimate for the spin-dependent tunnel matrix element $V_\text{s}$, the values quoted below for 'Spin relaxation (spin-orbit)' and '1-qubit gate (spin, $\pi/2$)' should not be regarded as predictions. (2) Relaxation refers to phonon-emission-mediated relaxation. (3) The $\pi/2$ single-qubit gate time is 1/4 times the inverse Rabi frequency. (4) 'On-donor flip-flop relaxation in bulk' is calculated assuming $1.2\,\text{K}$ and $h\times9\,\text{GHz}$ flip-flop splitting.}
\begin{tabular}{l  c  c}
Processes/gates & Rate & Time \\
\hline \hline
Flip-flop qubit gate times \\
\hline
1-qubit gate ($\pi/2$) & $22\,\text{MHz}$ & $45\,\text{ns}$ \\
2-qubit gate ($\sqrt{\text{SWAP}}$, cavity-mediated)\cite{Tosi} & $2.5\,\text{MHz}$ & $400\,\text{ns}$ \\
2-qubit gate ($\sqrt{\text{SWAP}}$, dipole-dipole)\cite{Tosi} & $25\,\text{MHz}$ & $40\,\text{ns}$ \\
\hline
Information loss of flip-flop qubit \\
\hline
Flip-flop relaxation & $3.7\,\text{kHz}$ & $270\,\mu\text{s}$ \\
Spin relaxation (spin-orbit, $\theta=\pi/2$) & $98\,\text{Hz}$ & $10\,\text{ms}$ \\
Spin relaxation ($g$-tensor mod., $\theta=\pi/4$) & $140\,\text{Hz}$ & $7\,\text{ms}$ \\
Electrically-induced dephasing \cite{Tosi} & $[1,300]\,\text{Hz}$ & $[3.3\,\text{ms},1\,\text{s}]$ \\
\hline
Further time scales\\
\hline
On-donor flip-flop relaxation in bulk\cite{Pines}  & $2.9\times10^{-5}\,\text{Hz}$ & $3.4\times10^4\,\text{s}$ \\
Orbital relaxation & $490\,\text{kHz}$ & $2.1\,\mu\text{s}$ \\
1-qubit gate (spin, $\pi/2$) & $3.6\,\text{MHz}$ & $280\,\text{ns}$ \\
\end{tabular}
\end{table}
\endgroup

\section{Conclusions}

We described fast, valley-enhanced relaxation 
mechanisms (orbital, flip-flop and electron spin relaxation)
for a gate-controlled P donor electron close to a
silicon/barrier interface. 
For the flip-flop qubit setup and the proposed qubit
working point, we have found
that  the flip-flop relaxation can be approximately 8 orders of 
magnitude faster than in bulk. 
The predicted relaxation time scale is $\sim 100 \, \mu \text{s}$,
still longer than 
the expected 
single-qubit ($40\,\text{ns}$) and two-qubit ($40-400\,\text{ns}$) gate times\cite{Tosi}.
Nevertheless, relaxation might dominate dephasing,
if our estimates as well as 
the inhomogeneous dephasing rate estimate\cite{Tosi}
$1/T_\text{2,ff}^* \sim [1,300]$ Hz are reliable. The relevant time scales are listed in Table \ref{tab:timescales}.

We also discussed analogous, valley-enhanced 
mechanisms inducing orbital and electron spin relaxation.
Since gate control of donor electrons near interfaces 
is an ubiquitous ingredient of donor-based quantum-computing 
schemes,
the relevance of the mechanisms described here extends
beyond the considered
specific flip-flop qubit architecture.

\appendix

\begin{acknowledgments}
We thank D.~Culcer, R.~Joynt, and R.~Rahman for useful
discussions, and 
M.~Calder\'on, A.~Morello, and G.~Tosi for
their helpful and constructive feedback on the manuscript.
We acknowledge funding from 
the EU Marie Curie Career Integration Grant CIG-293834,
Hungarian OTKA Grants No.~PD 100373 and 108676, 
the Gordon Godfrey Bequest, and the EU ERC Starting Grant 258789.
A.~P.~was supported by the 
J\'anos Bolyai Scholarship of the Hungarian Academy of Sciences.
\end{acknowledgments}

\bibliography{refs_Si}

\begin{thebibliography}{49}%
\makeatletter
\providecommand \@ifxundefined [1]{%
 \@ifx{#1\undefined}
}%
\providecommand \@ifnum [1]{%
 \ifnum #1\expandafter \@firstoftwo
 \else \expandafter \@secondoftwo
 \fi
}%
\providecommand \@ifx [1]{%
 \ifx #1\expandafter \@firstoftwo
 \else \expandafter \@secondoftwo
 \fi
}%
\providecommand \natexlab [1]{#1}%
\providecommand \enquote  [1]{``#1''}%
\providecommand \bibnamefont  [1]{#1}%
\providecommand \bibfnamefont [1]{#1}%
\providecommand \citenamefont [1]{#1}%
\providecommand \href@noop [0]{\@secondoftwo}%
\providecommand \href [0]{\begingroup \@sanitize@url \@href}%
\providecommand \@href[1]{\@@startlink{#1}\@@href}%
\providecommand \@@href[1]{\endgroup#1\@@endlink}%
\providecommand \@sanitize@url [0]{\catcode `\\12\catcode `\$12\catcode
  `\&12\catcode `\#12\catcode `\^12\catcode `\_12\catcode `\%12\relax}%
\providecommand \@@startlink[1]{}%
\providecommand \@@endlink[0]{}%
\providecommand \url  [0]{\begingroup\@sanitize@url \@url }%
\providecommand \@url [1]{\endgroup\@href {#1}{\urlprefix }}%
\providecommand \urlprefix  [0]{URL }%
\providecommand \Eprint [0]{\href }%
\providecommand \doibase [0]{http://dx.doi.org/}%
\providecommand \selectlanguage [0]{\@gobble}%
\providecommand \bibinfo  [0]{\@secondoftwo}%
\providecommand \bibfield  [0]{\@secondoftwo}%
\providecommand \translation [1]{[#1]}%
\providecommand \BibitemOpen [0]{}%
\providecommand \bibitemStop [0]{}%
\providecommand \bibitemNoStop [0]{.\EOS\space}%
\providecommand \EOS [0]{\spacefactor3000\relax}%
\providecommand \BibitemShut  [1]{\csname bibitem#1\endcsname}%
\let\auto@bib@innerbib\@empty
\bibitem [{\citenamefont {Kane}(1998)}]{Kane_Nature98}%
  \BibitemOpen
  \bibfield  {author} {\bibinfo {author} {\bibfnamefont {B.~E.}\ \bibnamefont
  {Kane}},\ }\href@noop {} {\bibfield  {journal} {\bibinfo  {journal} {Nature}\
  }\textbf {\bibinfo {volume} {393}},\ \bibinfo {pages} {133} (\bibinfo {year}
  {1998})}\BibitemShut {NoStop}%
\bibitem [{\citenamefont {Morton}\ \emph {et~al.}(2011)\citenamefont {Morton},
  \citenamefont {McCamey}, \citenamefont {Eriksson},\ and\ \citenamefont
  {Lyon}}]{Morton_Si_QC_QmLim_Nat11}%
  \BibitemOpen
  \bibfield  {author} {\bibinfo {author} {\bibfnamefont {J.}~\bibnamefont
  {Morton}}, \bibinfo {author} {\bibfnamefont {D.}~\bibnamefont {McCamey}},
  \bibinfo {author} {\bibfnamefont {M.}~\bibnamefont {Eriksson}}, \ and\
  \bibinfo {author} {\bibfnamefont {S.}~\bibnamefont {Lyon}},\ }\href@noop {}
  {\bibfield  {journal} {\bibinfo  {journal} {Nature}\ }\textbf {\bibinfo
  {volume} {479}},\ \bibinfo {pages} {345} (\bibinfo {year}
  {2011})}\BibitemShut {NoStop}%
\bibitem [{\citenamefont {Zwanenburg}\ \emph {et~al.}(2013)\citenamefont
  {Zwanenburg}, \citenamefont {Dzurak}, \citenamefont {Morello}, \citenamefont
  {Simmons}, \citenamefont {Hollenberg}, \citenamefont {Klimeck}, \citenamefont
  {Rogge}, \citenamefont {Coppersmith},\ and\ \citenamefont
  {Eriksson}}]{Zwanenburg_SiQmEl_RMP13}%
  \BibitemOpen
  \bibfield  {author} {\bibinfo {author} {\bibfnamefont {F.~A.}\ \bibnamefont
  {Zwanenburg}}, \bibinfo {author} {\bibfnamefont {A.~S.}\ \bibnamefont
  {Dzurak}}, \bibinfo {author} {\bibfnamefont {A.}~\bibnamefont {Morello}},
  \bibinfo {author} {\bibfnamefont {M.}~\bibnamefont {Simmons}}, \bibinfo
  {author} {\bibfnamefont {L.}~\bibnamefont {Hollenberg}}, \bibinfo {author}
  {\bibfnamefont {G.}~\bibnamefont {Klimeck}}, \bibinfo {author} {\bibfnamefont
  {S.}~\bibnamefont {Rogge}}, \bibinfo {author} {\bibfnamefont
  {S.}~\bibnamefont {Coppersmith}}, \ and\ \bibinfo {author} {\bibfnamefont
  {M.}~\bibnamefont {Eriksson}},\ }\href@noop {} {\bibfield  {journal}
  {\bibinfo  {journal} {Rev. Mod. Phys.}\ }\textbf {\bibinfo {volume} {85}},\
  \bibinfo {pages} {961} (\bibinfo {year} {2013})}\BibitemShut {NoStop}%
\bibitem [{\citenamefont {Feher}(1959)}]{Feher_PR59}%
  \BibitemOpen
  \bibfield  {author} {\bibinfo {author} {\bibfnamefont {G.}~\bibnamefont
  {Feher}},\ }\href@noop {} {\bibfield  {journal} {\bibinfo  {journal} {Phys.
  Rev.}\ }\textbf {\bibinfo {volume} {114}},\ \bibinfo {pages} {1219} (\bibinfo
  {year} {1959})}\BibitemShut {NoStop}%
\bibitem [{\citenamefont {Feher}\ and\ \citenamefont
  {Gere}(1959)}]{FeherGere_PR59}%
  \BibitemOpen
  \bibfield  {author} {\bibinfo {author} {\bibfnamefont {G.}~\bibnamefont
  {Feher}}\ and\ \bibinfo {author} {\bibfnamefont {E.~A.}\ \bibnamefont
  {Gere}},\ }\href@noop {} {\bibfield  {journal} {\bibinfo  {journal} {Phys.
  Rev.}\ }\textbf {\bibinfo {volume} {114}},\ \bibinfo {pages} {1245} (\bibinfo
  {year} {1959})}\BibitemShut {NoStop}%
\bibitem [{\citenamefont {Roth}(1960)}]{Roth_PR60}%
  \BibitemOpen
  \bibfield  {author} {\bibinfo {author} {\bibfnamefont {L.~M.}\ \bibnamefont
  {Roth}},\ }\href@noop {} {\bibfield  {journal} {\bibinfo  {journal} {Phys.\
  Rev.}\ }\textbf {\bibinfo {volume} {118}},\ \bibinfo {pages} {1534} (\bibinfo
  {year} {1960})}\BibitemShut {NoStop}%
\bibitem [{\citenamefont {Hasegawa}(1960)}]{Hasegawa_PR60}%
  \BibitemOpen
  \bibfield  {author} {\bibinfo {author} {\bibfnamefont {H.}~\bibnamefont
  {Hasegawa}},\ }\href@noop {} {\bibfield  {journal} {\bibinfo  {journal}
  {Phys.\ Rev.}\ }\textbf {\bibinfo {volume} {118}},\ \bibinfo {pages} {1523}
  (\bibinfo {year} {1960})}\BibitemShut {NoStop}%
\bibitem [{\citenamefont {Tahan}\ \emph {et~al.}(2002)\citenamefont {Tahan},
  \citenamefont {Friesen},\ and\ \citenamefont {Joynt}}]{Tahan2002}%
  \BibitemOpen
  \bibfield  {author} {\bibinfo {author} {\bibfnamefont {C.}~\bibnamefont
  {Tahan}}, \bibinfo {author} {\bibfnamefont {M.}~\bibnamefont {Friesen}}, \
  and\ \bibinfo {author} {\bibfnamefont {R.}~\bibnamefont {Joynt}},\ }\href
  {\doibase 10.1103/PhysRevB.66.035314} {\bibfield  {journal} {\bibinfo
  {journal} {Phys. Rev. B}\ }\textbf {\bibinfo {volume} {66}},\ \bibinfo
  {pages} {035314} (\bibinfo {year} {2002})}\BibitemShut {NoStop}%
\bibitem [{\citenamefont {Tyryshkin}\ \emph {et~al.}(2003)\citenamefont
  {Tyryshkin}, \citenamefont {Lyon}, \citenamefont {Astashkin},\ and\
  \citenamefont {Raitsimring}}]{Tyryshkin_PRB03}%
  \BibitemOpen
  \bibfield  {author} {\bibinfo {author} {\bibfnamefont {A.~M.}\ \bibnamefont
  {Tyryshkin}}, \bibinfo {author} {\bibfnamefont {S.~A.}\ \bibnamefont {Lyon}},
  \bibinfo {author} {\bibfnamefont {A.~V.}\ \bibnamefont {Astashkin}}, \ and\
  \bibinfo {author} {\bibfnamefont {A.~M.}\ \bibnamefont {Raitsimring}},\
  }\href@noop {} {\bibfield  {journal} {\bibinfo  {journal} {Phys. Rev. B}\
  }\textbf {\bibinfo {volume} {68}},\ \bibinfo {pages} {193207} (\bibinfo
  {year} {2003})}\BibitemShut {NoStop}%
\bibitem [{\citenamefont {Tyryshkin}\ \emph {et~al.}(2006)\citenamefont
  {Tyryshkin}, \citenamefont {Morton}, \citenamefont {Benjamin}, \citenamefont
  {Ardavan}, \citenamefont {Briggs}, \citenamefont {Ager},\ and\ \citenamefont
  {Lyon}}]{Tyryshkin_JPC06}%
  \BibitemOpen
  \bibfield  {author} {\bibinfo {author} {\bibfnamefont {A.~M.}\ \bibnamefont
  {Tyryshkin}}, \bibinfo {author} {\bibfnamefont {J.~J.~L.}\ \bibnamefont
  {Morton}}, \bibinfo {author} {\bibfnamefont {S.~C.}\ \bibnamefont
  {Benjamin}}, \bibinfo {author} {\bibfnamefont {A.}~\bibnamefont {Ardavan}},
  \bibinfo {author} {\bibfnamefont {G.~A.~D.}\ \bibnamefont {Briggs}}, \bibinfo
  {author} {\bibfnamefont {J.~W.}\ \bibnamefont {Ager}}, \ and\ \bibinfo
  {author} {\bibfnamefont {S.~A.}\ \bibnamefont {Lyon}},\ }\href@noop {}
  {\bibfield  {journal} {\bibinfo  {journal} {J. Phys. Condens. Matter}\
  }\textbf {\bibinfo {volume} {18}},\ \bibinfo {pages} {S783} (\bibinfo {year}
  {2006})}\BibitemShut {NoStop}%
\bibitem [{\citenamefont {Morello}\ \emph {et~al.}(2010)\citenamefont
  {Morello}, \citenamefont {Pla}, \citenamefont {Zwanenburg}, \citenamefont
  {Chan}, \citenamefont {Huebl}, \citenamefont {Mottonen}, \citenamefont
  {Nugroho}, \citenamefont {Yang}, \citenamefont {van Donkelaar}, \citenamefont
  {Alves}, \citenamefont {Jamieson}, \citenamefont {Escott}, \citenamefont
  {Hollenberg}, \citenamefont {Clark},\ and\ \citenamefont
  {Dzurak}}]{Morello_1glShot_Nature10}%
  \BibitemOpen
  \bibfield  {author} {\bibinfo {author} {\bibfnamefont {A.}~\bibnamefont
  {Morello}}, \bibinfo {author} {\bibfnamefont {J.~J.}\ \bibnamefont {Pla}},
  \bibinfo {author} {\bibfnamefont {F.~A.}\ \bibnamefont {Zwanenburg}},
  \bibinfo {author} {\bibfnamefont {K.~W.}\ \bibnamefont {Chan}}, \bibinfo
  {author} {\bibfnamefont {H.}~\bibnamefont {Huebl}}, \bibinfo {author}
  {\bibfnamefont {M.}~\bibnamefont {Mottonen}}, \bibinfo {author}
  {\bibfnamefont {C.~D.}\ \bibnamefont {Nugroho}}, \bibinfo {author}
  {\bibfnamefont {C.}~\bibnamefont {Yang}}, \bibinfo {author} {\bibfnamefont
  {J.~A.}\ \bibnamefont {van Donkelaar}}, \bibinfo {author} {\bibfnamefont
  {A.~D.~C.}\ \bibnamefont {Alves}}, \bibinfo {author} {\bibfnamefont {D.~N.}\
  \bibnamefont {Jamieson}}, \bibinfo {author} {\bibfnamefont {C.~C.}\
  \bibnamefont {Escott}}, \bibinfo {author} {\bibfnamefont {L.~C.~L.}\
  \bibnamefont {Hollenberg}}, \bibinfo {author} {\bibfnamefont {R.~G.}\
  \bibnamefont {Clark}}, \ and\ \bibinfo {author} {\bibfnamefont {A.~S.}\
  \bibnamefont {Dzurak}},\ }\href@noop {} {\bibfield  {journal} {\bibinfo
  {journal} {Nature}\ }\textbf {\bibinfo {volume} {467}},\ \bibinfo {pages}
  {687} (\bibinfo {year} {2010})}\BibitemShut {NoStop}%
\bibitem [{\citenamefont {Tyryshkin}\ \emph {et~al.}(2012)\citenamefont
  {Tyryshkin}, \citenamefont {Tojo}, \citenamefont {Morton}, \citenamefont
  {Riemann}, \citenamefont {Abrosimov}, \citenamefont {Becker}, \citenamefont
  {Pohl}, \citenamefont {Schenkel}, \citenamefont {Thewalt}, \citenamefont
  {Itoh},\ and\ \citenamefont {Lyon}}]{Tyryshkin_IsoPureSiDnr_T2Secs_11}%
  \BibitemOpen
  \bibfield  {author} {\bibinfo {author} {\bibfnamefont {A.~M.}\ \bibnamefont
  {Tyryshkin}}, \bibinfo {author} {\bibfnamefont {S.}~\bibnamefont {Tojo}},
  \bibinfo {author} {\bibfnamefont {J.~J.~L.}\ \bibnamefont {Morton}}, \bibinfo
  {author} {\bibfnamefont {H.}~\bibnamefont {Riemann}}, \bibinfo {author}
  {\bibfnamefont {N.~V.}\ \bibnamefont {Abrosimov}}, \bibinfo {author}
  {\bibfnamefont {P.}~\bibnamefont {Becker}}, \bibinfo {author} {\bibfnamefont
  {H.-J.}\ \bibnamefont {Pohl}}, \bibinfo {author} {\bibfnamefont
  {T.}~\bibnamefont {Schenkel}}, \bibinfo {author} {\bibfnamefont {M.~L.~W.}\
  \bibnamefont {Thewalt}}, \bibinfo {author} {\bibfnamefont {K.~M.}\
  \bibnamefont {Itoh}}, \ and\ \bibinfo {author} {\bibfnamefont {S.~A.}\
  \bibnamefont {Lyon}},\ }\href@noop {} {\bibfield  {journal} {\bibinfo
  {journal} {Nat Mater}\ }\textbf {\bibinfo {volume} {11}},\ \bibinfo {pages}
  {143} (\bibinfo {year} {2012})}\BibitemShut {NoStop}%
\bibitem [{\citenamefont {Tahan}\ and\ \citenamefont
  {Joynt}(2014)}]{Tahan2014}%
  \BibitemOpen
  \bibfield  {author} {\bibinfo {author} {\bibfnamefont {C.}~\bibnamefont
  {Tahan}}\ and\ \bibinfo {author} {\bibfnamefont {R.}~\bibnamefont {Joynt}},\
  }\href {\doibase 10.1103/PhysRevB.89.075302} {\bibfield  {journal} {\bibinfo
  {journal} {Phys. Rev. B}\ }\textbf {\bibinfo {volume} {89}},\ \bibinfo
  {pages} {075302} (\bibinfo {year} {2014})}\BibitemShut {NoStop}%
\bibitem [{\citenamefont {Pla}\ \emph {et~al.}(2012)\citenamefont {Pla},
  \citenamefont {Tan}, \citenamefont {Dehollain}, \citenamefont {Lim},
  \citenamefont {Morton}, \citenamefont {Jamieson}, \citenamefont {Dzurak},\
  and\ \citenamefont {Morello}}]{Pla-electronspin}%
  \BibitemOpen
  \bibfield  {author} {\bibinfo {author} {\bibfnamefont {J.~J.}\ \bibnamefont
  {Pla}}, \bibinfo {author} {\bibfnamefont {K.~Y.}\ \bibnamefont {Tan}},
  \bibinfo {author} {\bibfnamefont {J.~P.}\ \bibnamefont {Dehollain}}, \bibinfo
  {author} {\bibfnamefont {W.~H.}\ \bibnamefont {Lim}}, \bibinfo {author}
  {\bibfnamefont {J.~J.~L.}\ \bibnamefont {Morton}}, \bibinfo {author}
  {\bibfnamefont {D.~N.}\ \bibnamefont {Jamieson}}, \bibinfo {author}
  {\bibfnamefont {A.~S.}\ \bibnamefont {Dzurak}}, \ and\ \bibinfo {author}
  {\bibfnamefont {A.}~\bibnamefont {Morello}},\ }\href@noop {} {\bibfield
  {journal} {\bibinfo  {journal} {Nature}\ }\textbf {\bibinfo {volume} {489}},\
  \bibinfo {pages} {541} (\bibinfo {year} {2012})}\BibitemShut {NoStop}%
\bibitem [{\citenamefont {Pla}\ \emph {et~al.}(2013)\citenamefont {Pla},
  \citenamefont {Tan}, \citenamefont {Dehollain}, \citenamefont {Lim},
  \citenamefont {Morton}, \citenamefont {Zwanenburg}, \citenamefont {Jamieson},
  \citenamefont {Dzurak},\ and\ \citenamefont {Morello}}]{Pla-nuclearspin}%
  \BibitemOpen
  \bibfield  {author} {\bibinfo {author} {\bibfnamefont {J.~J.}\ \bibnamefont
  {Pla}}, \bibinfo {author} {\bibfnamefont {K.~Y.}\ \bibnamefont {Tan}},
  \bibinfo {author} {\bibfnamefont {J.~P.}\ \bibnamefont {Dehollain}}, \bibinfo
  {author} {\bibfnamefont {W.~H.}\ \bibnamefont {Lim}}, \bibinfo {author}
  {\bibfnamefont {J.~J.~L.}\ \bibnamefont {Morton}}, \bibinfo {author}
  {\bibfnamefont {F.~A.}\ \bibnamefont {Zwanenburg}}, \bibinfo {author}
  {\bibfnamefont {D.~N.}\ \bibnamefont {Jamieson}}, \bibinfo {author}
  {\bibfnamefont {A.~S.}\ \bibnamefont {Dzurak}}, \ and\ \bibinfo {author}
  {\bibfnamefont {A.}~\bibnamefont {Morello}},\ }\href@noop {} {\bibfield
  {journal} {\bibinfo  {journal} {Nature}\ }\textbf {\bibinfo {volume} {496}},\
  \bibinfo {pages} {334} (\bibinfo {year} {2013})}\BibitemShut {NoStop}%
\bibitem [{\citenamefont {Muhonen}\ \emph {et~al.}(2014)\citenamefont
  {Muhonen}, \citenamefont {Dehollain}, \citenamefont {Laucht}, \citenamefont
  {Hudson}, \citenamefont {Sekiguchi}, \citenamefont {Itoh}, \citenamefont
  {Jamieson}, \citenamefont {McCallum}, \citenamefont {Dzurak},\ and\
  \citenamefont {Morello}}]{Muhonen_Store_14}%
  \BibitemOpen
  \bibfield  {author} {\bibinfo {author} {\bibfnamefont {J.}~\bibnamefont
  {Muhonen}}, \bibinfo {author} {\bibfnamefont {J.}~\bibnamefont {Dehollain}},
  \bibinfo {author} {\bibfnamefont {A.}~\bibnamefont {Laucht}}, \bibinfo
  {author} {\bibfnamefont {F.}~\bibnamefont {Hudson}}, \bibinfo {author}
  {\bibfnamefont {T.}~\bibnamefont {Sekiguchi}}, \bibinfo {author}
  {\bibfnamefont {K.}~\bibnamefont {Itoh}}, \bibinfo {author} {\bibfnamefont
  {D.}~\bibnamefont {Jamieson}}, \bibinfo {author} {\bibfnamefont
  {J.}~\bibnamefont {McCallum}}, \bibinfo {author} {\bibfnamefont
  {A.}~\bibnamefont {Dzurak}}, \ and\ \bibinfo {author} {\bibfnamefont
  {A.}~\bibnamefont {Morello}},\ }\href@noop {} {\bibfield  {journal} {\bibinfo
   {journal} {Nature Nanotechnology}\ }\textbf {\bibinfo {volume} {9}},\
  \bibinfo {pages} {986 } (\bibinfo {year} {2014})}\BibitemShut {NoStop}%
\bibitem [{\citenamefont {Itoh}\ and\ \citenamefont
  {Watanabe}(2014)}]{Itoh_MRS}%
  \BibitemOpen
  \bibfield  {author} {\bibinfo {author} {\bibfnamefont {K.~M.}\ \bibnamefont
  {Itoh}}\ and\ \bibinfo {author} {\bibfnamefont {H.}~\bibnamefont
  {Watanabe}},\ }\href@noop {} {\bibfield  {journal} {\bibinfo  {journal} {MRS
  Communications}\ }\textbf {\bibinfo {volume} {4}},\ \bibinfo {pages} {143}
  (\bibinfo {year} {2014})}\BibitemShut {NoStop}%
\bibitem [{\citenamefont {Vrijen}\ \emph {et~al.}(2000)\citenamefont {Vrijen},
  \citenamefont {Yablonovitch}, \citenamefont {Wang}, \citenamefont {Jiang},
  \citenamefont {Balandin}, \citenamefont {Roychowdhury}, \citenamefont {Mor},\
  and\ \citenamefont {DiVincenzo}}]{Vrijen_PRA00}%
  \BibitemOpen
  \bibfield  {author} {\bibinfo {author} {\bibfnamefont {R.}~\bibnamefont
  {Vrijen}}, \bibinfo {author} {\bibfnamefont {E.}~\bibnamefont
  {Yablonovitch}}, \bibinfo {author} {\bibfnamefont {K.}~\bibnamefont {Wang}},
  \bibinfo {author} {\bibfnamefont {H.~W.}\ \bibnamefont {Jiang}}, \bibinfo
  {author} {\bibfnamefont {A.}~\bibnamefont {Balandin}}, \bibinfo {author}
  {\bibfnamefont {V.}~\bibnamefont {Roychowdhury}}, \bibinfo {author}
  {\bibfnamefont {T.}~\bibnamefont {Mor}}, \ and\ \bibinfo {author}
  {\bibfnamefont {D.}~\bibnamefont {DiVincenzo}},\ }\href@noop {} {\bibfield
  {journal} {\bibinfo  {journal} {Phys.\ Rev.\ A}\ }\textbf {\bibinfo {volume}
  {62}},\ \bibinfo {pages} {012306} (\bibinfo {year} {2000})}\BibitemShut
  {NoStop}%
\bibitem [{\citenamefont {Calder\'on}\ \emph {et~al.}(2006)\citenamefont
  {Calder\'on}, \citenamefont {Koiller}, \citenamefont {Hu},\ and\
  \citenamefont {Das~Sarma}}]{Calderon_quantumcontrol}%
  \BibitemOpen
  \bibfield  {author} {\bibinfo {author} {\bibfnamefont {M.~J.}\ \bibnamefont
  {Calder\'on}}, \bibinfo {author} {\bibfnamefont {B.}~\bibnamefont {Koiller}},
  \bibinfo {author} {\bibfnamefont {X.}~\bibnamefont {Hu}}, \ and\ \bibinfo
  {author} {\bibfnamefont {S.}~\bibnamefont {Das~Sarma}},\ }\href {\doibase
  10.1103/PhysRevLett.96.096802} {\bibfield  {journal} {\bibinfo  {journal}
  {Phys. Rev. Lett.}\ }\textbf {\bibinfo {volume} {96}},\ \bibinfo {pages}
  {096802} (\bibinfo {year} {2006})}\BibitemShut {NoStop}%
\bibitem [{\citenamefont {Calder\'{o}n}\ \emph {et~al.}(2008)\citenamefont
  {Calder\'{o}n}, \citenamefont {Koiller},\ and\ \citenamefont {{Das
  Sarma}}}]{Calderon_PRB08}%
  \BibitemOpen
  \bibfield  {author} {\bibinfo {author} {\bibfnamefont {M.~J.}\ \bibnamefont
  {Calder\'{o}n}}, \bibinfo {author} {\bibfnamefont {B.}~\bibnamefont
  {Koiller}}, \ and\ \bibinfo {author} {\bibfnamefont {S.}~\bibnamefont {{Das
  Sarma}}},\ }\href@noop {} {\bibfield  {journal} {\bibinfo  {journal} {Phys.\
  Rev.\ B}\ }\textbf {\bibinfo {volume} {77}},\ \bibinfo {pages} {155302}
  (\bibinfo {year} {2008})}\BibitemShut {NoStop}%
\bibitem [{\citenamefont {Lansbergen}\ \emph {et~al.}(2008)\citenamefont
  {Lansbergen}, \citenamefont {Rahman}, \citenamefont {Wellard}, \citenamefont
  {Woo}, \citenamefont {Caro}, \citenamefont {Collaert}, \citenamefont
  {Biesemans}, \citenamefont {Klimeck}, \citenamefont {Hollenberg},\ and\
  \citenamefont {Rogge}}]{Lansbergen_donorcontrol}%
  \BibitemOpen
  \bibfield  {author} {\bibinfo {author} {\bibfnamefont {G.~P.}\ \bibnamefont
  {Lansbergen}}, \bibinfo {author} {\bibfnamefont {R.}~\bibnamefont {Rahman}},
  \bibinfo {author} {\bibfnamefont {C.~J.}\ \bibnamefont {Wellard}}, \bibinfo
  {author} {\bibfnamefont {I.}~\bibnamefont {Woo}}, \bibinfo {author}
  {\bibfnamefont {J.}~\bibnamefont {Caro}}, \bibinfo {author} {\bibfnamefont
  {N.}~\bibnamefont {Collaert}}, \bibinfo {author} {\bibfnamefont
  {S.}~\bibnamefont {Biesemans}}, \bibinfo {author} {\bibfnamefont
  {G.}~\bibnamefont {Klimeck}}, \bibinfo {author} {\bibfnamefont {L.~C.~L.}\
  \bibnamefont {Hollenberg}}, \ and\ \bibinfo {author} {\bibfnamefont
  {S.}~\bibnamefont {Rogge}},\ }\href@noop {} {\bibfield  {journal} {\bibinfo
  {journal} {Nat Phys}\ }\textbf {\bibinfo {volume} {4}},\ \bibinfo {pages}
  {656} (\bibinfo {year} {2008})}\BibitemShut {NoStop}%
\bibitem [{\citenamefont {Rahman}\ \emph
  {et~al.}(2009{\natexlab{a}})\citenamefont {Rahman}, \citenamefont
  {Lansbergen}, \citenamefont {Park}, \citenamefont {Verduijn}, \citenamefont
  {Klimeck}, \citenamefont {Rogge},\ and\ \citenamefont
  {Hollenberg}}]{Rahman-orbitalstark}%
  \BibitemOpen
  \bibfield  {author} {\bibinfo {author} {\bibfnamefont {R.}~\bibnamefont
  {Rahman}}, \bibinfo {author} {\bibfnamefont {G.~P.}\ \bibnamefont
  {Lansbergen}}, \bibinfo {author} {\bibfnamefont {S.~H.}\ \bibnamefont
  {Park}}, \bibinfo {author} {\bibfnamefont {J.}~\bibnamefont {Verduijn}},
  \bibinfo {author} {\bibfnamefont {G.}~\bibnamefont {Klimeck}}, \bibinfo
  {author} {\bibfnamefont {S.}~\bibnamefont {Rogge}}, \ and\ \bibinfo {author}
  {\bibfnamefont {L.~C.~L.}\ \bibnamefont {Hollenberg}},\ }\href {\doibase
  10.1103/PhysRevB.80.165314} {\bibfield  {journal} {\bibinfo  {journal} {Phys.
  Rev. B}\ }\textbf {\bibinfo {volume} {80}},\ \bibinfo {pages} {165314}
  (\bibinfo {year} {2009}{\natexlab{a}})}\BibitemShut {NoStop}%
\bibitem [{\citenamefont {Baena}\ \emph {et~al.}(2012)\citenamefont {Baena},
  \citenamefont {Saraiva}, \citenamefont {Koiller},\ and\ \citenamefont
  {Calder\'on}}]{Baena}%
  \BibitemOpen
  \bibfield  {author} {\bibinfo {author} {\bibfnamefont {A.}~\bibnamefont
  {Baena}}, \bibinfo {author} {\bibfnamefont {A.~L.}\ \bibnamefont {Saraiva}},
  \bibinfo {author} {\bibfnamefont {B.}~\bibnamefont {Koiller}}, \ and\
  \bibinfo {author} {\bibfnamefont {M.~J.}\ \bibnamefont {Calder\'on}},\ }\href
  {\doibase 10.1103/PhysRevB.86.035317} {\bibfield  {journal} {\bibinfo
  {journal} {Phys. Rev. B}\ }\textbf {\bibinfo {volume} {86}},\ \bibinfo
  {pages} {035317} (\bibinfo {year} {2012})}\BibitemShut {NoStop}%
\bibitem [{\citenamefont {Laucht}\ \emph {et~al.}(2015)\citenamefont {Laucht},
  \citenamefont {Muhonen}, \citenamefont {Mohiyaddin}, \citenamefont {Kalra},
  \citenamefont {Dehollain}, \citenamefont {Freer}, \citenamefont {Hudson},
  \citenamefont {Veldhorst}, \citenamefont {Rahman}, \citenamefont {Klimeck},
  \citenamefont {Itoh}, \citenamefont {Jamieson}, \citenamefont {McCallum},
  \citenamefont {Dzurak},\ and\ \citenamefont {Morello}}]{Laucht}%
  \BibitemOpen
  \bibfield  {author} {\bibinfo {author} {\bibfnamefont {A.}~\bibnamefont
  {Laucht}}, \bibinfo {author} {\bibfnamefont {J.~T.}\ \bibnamefont {Muhonen}},
  \bibinfo {author} {\bibfnamefont {F.~A.}\ \bibnamefont {Mohiyaddin}},
  \bibinfo {author} {\bibfnamefont {R.}~\bibnamefont {Kalra}}, \bibinfo
  {author} {\bibfnamefont {J.~P.}\ \bibnamefont {Dehollain}}, \bibinfo {author}
  {\bibfnamefont {S.}~\bibnamefont {Freer}}, \bibinfo {author} {\bibfnamefont
  {F.~E.}\ \bibnamefont {Hudson}}, \bibinfo {author} {\bibfnamefont
  {M.}~\bibnamefont {Veldhorst}}, \bibinfo {author} {\bibfnamefont
  {R.}~\bibnamefont {Rahman}}, \bibinfo {author} {\bibfnamefont
  {G.}~\bibnamefont {Klimeck}}, \bibinfo {author} {\bibfnamefont {K.~M.}\
  \bibnamefont {Itoh}}, \bibinfo {author} {\bibfnamefont {D.~N.}\ \bibnamefont
  {Jamieson}}, \bibinfo {author} {\bibfnamefont {J.~C.}\ \bibnamefont
  {McCallum}}, \bibinfo {author} {\bibfnamefont {A.~S.}\ \bibnamefont
  {Dzurak}}, \ and\ \bibinfo {author} {\bibfnamefont {A.}~\bibnamefont
  {Morello}},\ }\href {\doibase 10.1126/sciadv.1500022} {\bibfield  {journal}
  {\bibinfo  {journal} {Science Advances}\ }\textbf {\bibinfo {volume} {1}}
  (\bibinfo {year} {2015}),\ 10.1126/sciadv.1500022}\BibitemShut {NoStop}%
\bibitem [{\citenamefont {Tosi}\ \emph {et~al.}()\citenamefont {Tosi},
  \citenamefont {Mohiyaddin}, \citenamefont {Tenberg}, \citenamefont {Rahman},
  \citenamefont {Klimeck},\ and\ \citenamefont {Morello}}]{Tosi}%
  \BibitemOpen
  \bibfield  {author} {\bibinfo {author} {\bibfnamefont {G.}~\bibnamefont
  {Tosi}}, \bibinfo {author} {\bibfnamefont {F.~A.}\ \bibnamefont
  {Mohiyaddin}}, \bibinfo {author} {\bibfnamefont {S.~B.}\ \bibnamefont
  {Tenberg}}, \bibinfo {author} {\bibfnamefont {R.}~\bibnamefont {Rahman}},
  \bibinfo {author} {\bibfnamefont {G.}~\bibnamefont {Klimeck}}, \ and\
  \bibinfo {author} {\bibfnamefont {A.}~\bibnamefont {Morello}},\ }\href@noop
  {} {}\bibinfo {note} {ArXiv:1509.08538v1}\BibitemShut {NoStop}%
\bibitem [{\citenamefont {Urdampilleta}\ \emph {et~al.}(2015)\citenamefont
  {Urdampilleta}, \citenamefont {Chatterjee}, \citenamefont {Lo}, \citenamefont
  {Kobayashi}, \citenamefont {Mansir}, \citenamefont {Barraud}, \citenamefont
  {Betz}, \citenamefont {Rogge}, \citenamefont {Gonzalez-Zalba},\ and\
  \citenamefont {Morton}}]{Urdampilleta}%
  \BibitemOpen
  \bibfield  {author} {\bibinfo {author} {\bibfnamefont {M.}~\bibnamefont
  {Urdampilleta}}, \bibinfo {author} {\bibfnamefont {A.}~\bibnamefont
  {Chatterjee}}, \bibinfo {author} {\bibfnamefont {C.~C.}\ \bibnamefont {Lo}},
  \bibinfo {author} {\bibfnamefont {T.}~\bibnamefont {Kobayashi}}, \bibinfo
  {author} {\bibfnamefont {J.}~\bibnamefont {Mansir}}, \bibinfo {author}
  {\bibfnamefont {S.}~\bibnamefont {Barraud}}, \bibinfo {author} {\bibfnamefont
  {A.~C.}\ \bibnamefont {Betz}}, \bibinfo {author} {\bibfnamefont
  {S.}~\bibnamefont {Rogge}}, \bibinfo {author} {\bibfnamefont {M.~F.}\
  \bibnamefont {Gonzalez-Zalba}}, \ and\ \bibinfo {author} {\bibfnamefont
  {J.~J.~L.}\ \bibnamefont {Morton}},\ }\href {\doibase
  10.1103/PhysRevX.5.031024} {\bibfield  {journal} {\bibinfo  {journal} {Phys.
  Rev. X}\ }\textbf {\bibinfo {volume} {5}},\ \bibinfo {pages} {031024}
  (\bibinfo {year} {2015})}\BibitemShut {NoStop}%
\bibitem [{\citenamefont {Harvey-Collard}\ \emph {et~al.}()\citenamefont
  {Harvey-Collard}, \citenamefont {Jacobson}, \citenamefont {Rudolph},
  \citenamefont {Dominguez}, \citenamefont {Eyck}, \citenamefont {Wendt},
  \citenamefont {Pluym}, \citenamefont {Gamble}, \citenamefont {Lilly},
  \citenamefont {Pioro-Ladri\`ere},\ and\ \citenamefont
  {Carroll}}]{HarveyCollard}%
  \BibitemOpen
  \bibfield  {author} {\bibinfo {author} {\bibfnamefont {P.}~\bibnamefont
  {Harvey-Collard}}, \bibinfo {author} {\bibfnamefont {N.~T.}\ \bibnamefont
  {Jacobson}}, \bibinfo {author} {\bibfnamefont {M.}~\bibnamefont {Rudolph}},
  \bibinfo {author} {\bibfnamefont {J.}~\bibnamefont {Dominguez}}, \bibinfo
  {author} {\bibfnamefont {G.~A.~T.}\ \bibnamefont {Eyck}}, \bibinfo {author}
  {\bibfnamefont {J.~R.}\ \bibnamefont {Wendt}}, \bibinfo {author}
  {\bibfnamefont {T.}~\bibnamefont {Pluym}}, \bibinfo {author} {\bibfnamefont
  {J.~K.}\ \bibnamefont {Gamble}}, \bibinfo {author} {\bibfnamefont {M.~P.}\
  \bibnamefont {Lilly}}, \bibinfo {author} {\bibfnamefont {M.}~\bibnamefont
  {Pioro-Ladri\`ere}}, \ and\ \bibinfo {author} {\bibfnamefont {M.~S.}\
  \bibnamefont {Carroll}},\ }\href@noop {} {}\bibinfo {note}
  {ArXiv:1512.01606v1}\BibitemShut {NoStop}%
\bibitem [{\citenamefont {Golovach}\ \emph {et~al.}(2006)\citenamefont
  {Golovach}, \citenamefont {Borhani},\ and\ \citenamefont
  {Loss}}]{Golovach_EDSR_PRB06}%
  \BibitemOpen
  \bibfield  {author} {\bibinfo {author} {\bibfnamefont {V.~N.}\ \bibnamefont
  {Golovach}}, \bibinfo {author} {\bibfnamefont {M.}~\bibnamefont {Borhani}}, \
  and\ \bibinfo {author} {\bibfnamefont {D.}~\bibnamefont {Loss}},\ }\href@noop
  {} {\bibfield  {journal} {\bibinfo  {journal} {Phys.\ Rev.\ B}\ }\textbf
  {\bibinfo {volume} {74}},\ \bibinfo {pages} {165319} (\bibinfo {year}
  {2006})}\BibitemShut {NoStop}%
\bibitem [{\citenamefont {Flindt}\ \emph {et~al.}(2006)\citenamefont {Flindt},
  \citenamefont {S\o{}rensen},\ and\ \citenamefont {Flensberg}}]{Flindt}%
  \BibitemOpen
  \bibfield  {author} {\bibinfo {author} {\bibfnamefont {C.}~\bibnamefont
  {Flindt}}, \bibinfo {author} {\bibfnamefont {A.~S.}\ \bibnamefont
  {S\o{}rensen}}, \ and\ \bibinfo {author} {\bibfnamefont {K.}~\bibnamefont
  {Flensberg}},\ }\href {\doibase 10.1103/PhysRevLett.97.240501} {\bibfield
  {journal} {\bibinfo  {journal} {Phys. Rev. Lett.}\ }\textbf {\bibinfo
  {volume} {97}},\ \bibinfo {pages} {240501} (\bibinfo {year}
  {2006})}\BibitemShut {NoStop}%
\bibitem [{\citenamefont {Nowack}\ \emph {et~al.}(2007)\citenamefont {Nowack},
  \citenamefont {Koppens}, \citenamefont {Nazarov},\ and\ \citenamefont
  {Vandersypen}}]{Nowack_Science07}%
  \BibitemOpen
  \bibfield  {author} {\bibinfo {author} {\bibfnamefont {K.~C.}\ \bibnamefont
  {Nowack}}, \bibinfo {author} {\bibfnamefont {F.~H.~L.}\ \bibnamefont
  {Koppens}}, \bibinfo {author} {\bibfnamefont {Y.~V.}\ \bibnamefont
  {Nazarov}}, \ and\ \bibinfo {author} {\bibfnamefont {L.~M.~K.}\ \bibnamefont
  {Vandersypen}},\ }\href@noop {} {\bibfield  {journal} {\bibinfo  {journal}
  {Science}\ }\textbf {\bibinfo {volume} {318}},\ \bibinfo {pages} {1430}
  (\bibinfo {year} {2007})}\BibitemShut {NoStop}%
\bibitem [{\citenamefont {Blais}\ \emph {et~al.}(2004)\citenamefont {Blais},
  \citenamefont {Huang}, \citenamefont {Wallraff}, \citenamefont {Girvin},\
  and\ \citenamefont {Schoelkopf}}]{Blais}%
  \BibitemOpen
  \bibfield  {author} {\bibinfo {author} {\bibfnamefont {A.}~\bibnamefont
  {Blais}}, \bibinfo {author} {\bibfnamefont {R.-S.}\ \bibnamefont {Huang}},
  \bibinfo {author} {\bibfnamefont {A.}~\bibnamefont {Wallraff}}, \bibinfo
  {author} {\bibfnamefont {S.~M.}\ \bibnamefont {Girvin}}, \ and\ \bibinfo
  {author} {\bibfnamefont {R.~J.}\ \bibnamefont {Schoelkopf}},\ }\href
  {\doibase 10.1103/PhysRevA.69.062320} {\bibfield  {journal} {\bibinfo
  {journal} {Phys. Rev. A}\ }\textbf {\bibinfo {volume} {69}},\ \bibinfo
  {pages} {062320} (\bibinfo {year} {2004})}\BibitemShut {NoStop}%
\bibitem [{\citenamefont {Trif}\ \emph {et~al.}(2007)\citenamefont {Trif},
  \citenamefont {Golovach},\ and\ \citenamefont
  {Loss}}]{Trif_spinspincoupling}%
  \BibitemOpen
  \bibfield  {author} {\bibinfo {author} {\bibfnamefont {M.}~\bibnamefont
  {Trif}}, \bibinfo {author} {\bibfnamefont {V.~N.}\ \bibnamefont {Golovach}},
  \ and\ \bibinfo {author} {\bibfnamefont {D.}~\bibnamefont {Loss}},\ }\href
  {\doibase 10.1103/PhysRevB.75.085307} {\bibfield  {journal} {\bibinfo
  {journal} {Phys. Rev. B}\ }\textbf {\bibinfo {volume} {75}},\ \bibinfo
  {pages} {085307} (\bibinfo {year} {2007})}\BibitemShut {NoStop}%
\bibitem [{\citenamefont {Salfi}\ \emph {et~al.}(2015)\citenamefont {Salfi},
  \citenamefont {Mol}, \citenamefont {Culcer},\ and\ \citenamefont
  {Rogge}}]{Salfi_acceptorqubit}%
  \BibitemOpen
  \bibfield  {author} {\bibinfo {author} {\bibfnamefont {J.}~\bibnamefont
  {Salfi}}, \bibinfo {author} {\bibfnamefont {J.~A.}\ \bibnamefont {Mol}},
  \bibinfo {author} {\bibfnamefont {D.}~\bibnamefont {Culcer}}, \ and\ \bibinfo
  {author} {\bibfnamefont {S.}~\bibnamefont {Rogge}},\ }\href@noop {}
  {\bibfield  {journal} {\bibinfo  {journal} {arXiv:1508.04259}\ } (\bibinfo
  {year} {2015})}\BibitemShut {NoStop}%
\bibitem [{\citenamefont {Cullis}\ and\ \citenamefont {Marko}(1970)}]{Cullis}%
  \BibitemOpen
  \bibfield  {author} {\bibinfo {author} {\bibfnamefont {P.~R.}\ \bibnamefont
  {Cullis}}\ and\ \bibinfo {author} {\bibfnamefont {J.~R.}\ \bibnamefont
  {Marko}},\ }\href {\doibase 10.1103/PhysRevB.1.632} {\bibfield  {journal}
  {\bibinfo  {journal} {Phys. Rev. B}\ }\textbf {\bibinfo {volume} {1}},\
  \bibinfo {pages} {632} (\bibinfo {year} {1970})}\BibitemShut {NoStop}%
\bibitem [{\citenamefont {Koiller}\ \emph {et~al.}(2001)\citenamefont
  {Koiller}, \citenamefont {Hu},\ and\ \citenamefont {{Das
  Sarma}}}]{Koiller_PRL01}%
  \BibitemOpen
  \bibfield  {author} {\bibinfo {author} {\bibfnamefont {B.}~\bibnamefont
  {Koiller}}, \bibinfo {author} {\bibfnamefont {X.}~\bibnamefont {Hu}}, \ and\
  \bibinfo {author} {\bibfnamefont {S.}~\bibnamefont {{Das Sarma}}},\
  }\href@noop {} {\bibfield  {journal} {\bibinfo  {journal} {Phys. Rev. Lett.}\
  }\textbf {\bibinfo {volume} {88}},\ \bibinfo {pages} {027903} (\bibinfo
  {year} {2001})}\BibitemShut {NoStop}%
\bibitem [{\citenamefont {Pines}\ \emph {et~al.}(1957)\citenamefont {Pines},
  \citenamefont {Bardeen},\ and\ \citenamefont {Slichter}}]{Pines}%
  \BibitemOpen
  \bibfield  {author} {\bibinfo {author} {\bibfnamefont {D.}~\bibnamefont
  {Pines}}, \bibinfo {author} {\bibfnamefont {J.}~\bibnamefont {Bardeen}}, \
  and\ \bibinfo {author} {\bibfnamefont {C.~P.}\ \bibnamefont {Slichter}},\
  }\href {\doibase 10.1103/PhysRev.106.489} {\bibfield  {journal} {\bibinfo
  {journal} {Phys. Rev.}\ }\textbf {\bibinfo {volume} {106}},\ \bibinfo {pages}
  {489} (\bibinfo {year} {1957})}\BibitemShut {NoStop}%
\bibitem [{\citenamefont {Abragam}(1961)}]{Abragam}%
  \BibitemOpen
  \bibfield  {author} {\bibinfo {author} {\bibfnamefont {A.}~\bibnamefont
  {Abragam}},\ }\href@noop {} {\emph {\bibinfo {title} {The principles of
  nuclear magnetism}}}\ (\bibinfo  {publisher} {Oxford University Press},\
  \bibinfo {year} {1961})\BibitemShut {NoStop}%
\bibitem [{\citenamefont {Khaetskii}(2001)}]{Khaetskii-physicae}%
  \BibitemOpen
  \bibfield  {author} {\bibinfo {author} {\bibfnamefont {A.~V.}\ \bibnamefont
  {Khaetskii}},\ }\href@noop {} {\bibfield  {journal} {\bibinfo  {journal}
  {Physica E}\ }\textbf {\bibinfo {volume} {10}},\ \bibinfo {pages} {27}
  (\bibinfo {year} {2001})}\BibitemShut {NoStop}%
\bibitem [{\citenamefont {Erlingsson}\ and\ \citenamefont
  {Nazarov}(2002)}]{Erlingsson_2002_hyperfine}%
  \BibitemOpen
  \bibfield  {author} {\bibinfo {author} {\bibfnamefont {S.~I.}\ \bibnamefont
  {Erlingsson}}\ and\ \bibinfo {author} {\bibfnamefont {Y.~V.}\ \bibnamefont
  {Nazarov}},\ }\href@noop {} {\bibfield  {journal} {\bibinfo  {journal}
  {Phys.\ Rev.\ B}\ }\textbf {\bibinfo {volume} {66}},\ \bibinfo {pages}
  {155327} (\bibinfo {year} {2002})}\BibitemShut {NoStop}%
\bibitem [{\citenamefont {Rahman}\ \emph
  {et~al.}(2009{\natexlab{b}})\citenamefont {Rahman}, \citenamefont {Park},
  \citenamefont {Boykin}, \citenamefont {Klimeck}, \citenamefont {Rogge},\ and\
  \citenamefont {Hollenberg}}]{Rahman_gfactor}%
  \BibitemOpen
  \bibfield  {author} {\bibinfo {author} {\bibfnamefont {R.}~\bibnamefont
  {Rahman}}, \bibinfo {author} {\bibfnamefont {S.~H.}\ \bibnamefont {Park}},
  \bibinfo {author} {\bibfnamefont {T.~B.}\ \bibnamefont {Boykin}}, \bibinfo
  {author} {\bibfnamefont {G.}~\bibnamefont {Klimeck}}, \bibinfo {author}
  {\bibfnamefont {S.}~\bibnamefont {Rogge}}, \ and\ \bibinfo {author}
  {\bibfnamefont {L.~C.~L.}\ \bibnamefont {Hollenberg}},\ }\href {\doibase
  10.1103/PhysRevB.80.155301} {\bibfield  {journal} {\bibinfo  {journal} {Phys.
  Rev. B}\ }\textbf {\bibinfo {volume} {80}},\ \bibinfo {pages} {155301}
  (\bibinfo {year} {2009}{\natexlab{b}})}\BibitemShut {NoStop}%
\bibitem [{\citenamefont {Herring}\ and\ \citenamefont {Vogt}(1956)}]{Herring}%
  \BibitemOpen
  \bibfield  {author} {\bibinfo {author} {\bibfnamefont {C.}~\bibnamefont
  {Herring}}\ and\ \bibinfo {author} {\bibfnamefont {E.}~\bibnamefont {Vogt}},\
  }\href {\doibase 10.1103/PhysRev.101.944} {\bibfield  {journal} {\bibinfo
  {journal} {Phys. Rev.}\ }\textbf {\bibinfo {volume} {101}},\ \bibinfo {pages}
  {944} (\bibinfo {year} {1956})}\BibitemShut {NoStop}%
\bibitem [{\citenamefont {Yu}\ and\ \citenamefont
  {Cardona}(2010)}]{Yu_Cardona}%
  \BibitemOpen
  \bibfield  {author} {\bibinfo {author} {\bibfnamefont {P.~Y.}\ \bibnamefont
  {Yu}}\ and\ \bibinfo {author} {\bibfnamefont {M.}~\bibnamefont {Cardona}},\
  }\href@noop {} {\emph {\bibinfo {title} {Fundamentals of Semiconductors}}}\
  (\bibinfo  {publisher} {Springer},\ \bibinfo {address} {Berlin},\ \bibinfo
  {year} {2010})\BibitemShut {NoStop}%
\bibitem [{\citenamefont {Kohn}\ and\ \citenamefont
  {Luttinger}(1955)}]{Kohn_donorstates}%
  \BibitemOpen
  \bibfield  {author} {\bibinfo {author} {\bibfnamefont {W.}~\bibnamefont
  {Kohn}}\ and\ \bibinfo {author} {\bibfnamefont {J.~M.}\ \bibnamefont
  {Luttinger}},\ }\href {\doibase 10.1103/PhysRev.98.915} {\bibfield  {journal}
  {\bibinfo  {journal} {Phys. Rev.}\ }\textbf {\bibinfo {volume} {98}},\
  \bibinfo {pages} {915} (\bibinfo {year} {1955})}\BibitemShut {NoStop}%
\bibitem [{\citenamefont {Danon}\ and\ \citenamefont
  {Nazarov}(2009)}]{Danon_spinblockade}%
  \BibitemOpen
  \bibfield  {author} {\bibinfo {author} {\bibfnamefont {J.}~\bibnamefont
  {Danon}}\ and\ \bibinfo {author} {\bibfnamefont {Y.~V.}\ \bibnamefont
  {Nazarov}},\ }\href {\doibase 10.1103/PhysRevB.80.041301} {\bibfield
  {journal} {\bibinfo  {journal} {Phys. Rev. B}\ }\textbf {\bibinfo {volume}
  {80}},\ \bibinfo {pages} {041301} (\bibinfo {year} {2009})}\BibitemShut
  {NoStop}%
\bibitem [{\citenamefont {van Vleck}(1940)}]{vanVleck-cancellation}%
  \BibitemOpen
  \bibfield  {author} {\bibinfo {author} {\bibfnamefont {J.~H.}\ \bibnamefont
  {van Vleck}},\ }\href@noop {} {\bibfield  {journal} {\bibinfo  {journal}
  {Phys. Rev.}\ }\textbf {\bibinfo {volume} {57}},\ \bibinfo {pages} {426}
  (\bibinfo {year} {1940})}\BibitemShut {NoStop}%
\bibitem [{\citenamefont {Khaetskii}\ and\ \citenamefont
  {Nazarov}(2001)}]{Khaetskii_2001}%
  \BibitemOpen
  \bibfield  {author} {\bibinfo {author} {\bibfnamefont {A.}~\bibnamefont
  {Khaetskii}}\ and\ \bibinfo {author} {\bibfnamefont {Y.}~\bibnamefont
  {Nazarov}},\ }\href@noop {} {\bibfield  {journal} {\bibinfo  {journal}
  {Phys.\ Rev.\ B}\ }\textbf {\bibinfo {volume} {64}},\ \bibinfo {pages}
  {125316} (\bibinfo {year} {2001})}\BibitemShut {NoStop}%
\bibitem [{\citenamefont {Koiller}\ \emph {et~al.}(2002)\citenamefont
  {Koiller}, \citenamefont {Hu},\ and\ \citenamefont {{Das
  Sarma}}}]{Koiller_PRB02}%
  \BibitemOpen
  \bibfield  {author} {\bibinfo {author} {\bibfnamefont {B.}~\bibnamefont
  {Koiller}}, \bibinfo {author} {\bibfnamefont {X.}~\bibnamefont {Hu}}, \ and\
  \bibinfo {author} {\bibfnamefont {S.}~\bibnamefont {{Das Sarma}}},\
  }\href@noop {} {\bibfield  {journal} {\bibinfo  {journal} {Phys. Rev. B}\
  }\textbf {\bibinfo {volume} {66}},\ \bibinfo {pages} {115201} (\bibinfo
  {year} {2002})}\BibitemShut {NoStop}%
\bibitem [{\citenamefont {Salfi}\ \emph {et~al.}(2014)\citenamefont {Salfi},
  \citenamefont {Mol}, \citenamefont {Rahman}, \citenamefont {Klimeck},
  \citenamefont {Simmons}, \citenamefont {Hollenberg},\ and\ \citenamefont
  {Rogge}}]{Salfi_VlyInt_NM14}%
  \BibitemOpen
  \bibfield  {author} {\bibinfo {author} {\bibfnamefont {J.}~\bibnamefont
  {Salfi}}, \bibinfo {author} {\bibfnamefont {J.~A.}\ \bibnamefont {Mol}},
  \bibinfo {author} {\bibfnamefont {R.}~\bibnamefont {Rahman}}, \bibinfo
  {author} {\bibfnamefont {G.}~\bibnamefont {Klimeck}}, \bibinfo {author}
  {\bibfnamefont {M.~Y.}\ \bibnamefont {Simmons}}, \bibinfo {author}
  {\bibfnamefont {L.~C.~L.}\ \bibnamefont {Hollenberg}}, \ and\ \bibinfo
  {author} {\bibfnamefont {S.}~\bibnamefont {Rogge}},\ }\href@noop {}
  {\bibfield  {journal} {\bibinfo  {journal} {Nat.\ Mater.}\ }\textbf {\bibinfo
  {volume} {13}},\ \bibinfo {pages} {605} (\bibinfo {year} {2014})}\BibitemShut
  {NoStop}%
\bibitem [{\citenamefont {Friesen}(2005)}]{Friesen-starkeffect}%
  \BibitemOpen
  \bibfield  {author} {\bibinfo {author} {\bibfnamefont {M.}~\bibnamefont
  {Friesen}},\ }\href {\doibase 10.1103/PhysRevLett.94.186403} {\bibfield
  {journal} {\bibinfo  {journal} {Phys. Rev. Lett.}\ }\textbf {\bibinfo
  {volume} {94}},\ \bibinfo {pages} {186403} (\bibinfo {year}
  {2005})}\BibitemShut {NoStop}%
\end{thebibliography}%

\end{document}